\documentclass[aps,pra,preprint,showpacs,amsmath,amssymb]{revtex4}
\input{epsf}

\usepackage{graphicx}
\usepackage{array}
\usepackage{longtable}
\usepackage{rotating,booktabs}
\usepackage{booktabs,threeparttable}
\usepackage{amsmath}
\allowdisplaybreaks[4]

\begin{document}

\newcommand{\ket}[1]{\ensuremath{\left|#1\right\rangle}} 
\newcommand{\bra}[1]{\ensuremath{\left\langle #1 \right|}} 
\newcommand{\braket}[2]{\ensuremath{\left\langle #1 | #2 \right\rangle}} 

\title{Calculations of long-range three-body interactions for Li($2\,^2S$)-Li($2\,^2S$)-Li($2\,^2P$)}

\author{Pei-Gen Yan$^{1,2}$, Li-Yan Tang$^{2,*}$~\footnotetext{*Email: lytang@wipm.ac.cn}, Zong-Chao Yan$^{1,2,3}$, and James F. Babb$^{4}$}

\affiliation {$^1$Department of Physics, University of New Brunswick, Fredericton, New Brunswick, E3B 5A3, Canada\\
$^2$State Key Laboratory of Magnetic Resonance and Atomic and Molecular Physics, Wuhan Institute of Physics and Mathematics, Chinese Academy of Sciences, Wuhan 430071, People's Republic of China\\
$^3$Center for Cold Atom Physics, Chinese Academy of Sciences, Wuhan 430071, People's Republic of China\\
$^4$ITAMP, Harvard-Smithsonian Center for Astrophysics, Cambridge, Massachusetts 02138, USA}

\date{\today}

\begin{abstract}
General formulas for calculating the
several leading
long-range interactions among three identical atoms where two atoms are in identical $S$ states and the other atom is in a $P$ state are
obtained using perturbation theory for the energies up to second order.
The first order (dipolar)
interactions  depend on the geometrical configurations of the three atoms.
In second order, additive and nonadditive dispersion interactions are obtained.
The nonadditive interactions depend on the geometrical configurations
in marked contrast
to the case where all three atoms are in identical $S$ states, for which the
nonadditive (also known as triple-dipole
or as Axilrod-Muto-Teller) dispersion interactions appear at the third order.
The formalism is demonstrated by the calculation of the coefficients
for the  Li($2\,^2S$)-Li($2\,^2S$)-Li($2\,^{2}P$) system using
variationally-generated atomic lithium wave functions in Hylleraas coordinates.
The present dipolar coefficients and additive and nonadditive dispersion  coefficients
may be useful in constructing  precise potential energy surfaces
for this three lithium atom system.
\end{abstract}

\pacs{34.20.Cf, 32.10.Dk, 34.50.Dy}
\maketitle

\section{Introduction}\label{Int}

A considerable number of studies are devoted to
investigations of
the  long-range interactions between three ground state atoms~\cite{axilrod43,Axi51,bell70,Sto71,MosWorJez95,McDKumMea96,LiHun96,Marinescu97,yan97,tang12},
but the  case of three atoms with one atom in an excited state is less studied.
Reliable determinations of long-range interactions for the case of one atom in
an excited state might be a consideration
for characterizing excited trimers in photoassociation~\cite{JonTieLet06,SafNaiSte12,ZhaLiZha14},  in
implementing quantum information processing
with blockade mechanisms~\cite{PohBer09,BarRavLab14},
and in spectroscopic studies of highly excited bound trimer states~\cite{SheGaoSen07,LiuStaRos09,NolCot13,KifLiJak13}.

In the present work we use perturbation theory up to second order to derive general formulas for calculating the long-range interaction coefficients for  three like  atoms with two atoms in identical $S$ states and the other atom in an excited $P$ state.
We exhibit the additive ``dipolar'' (or dipole-dipole) interactions and additive dispersion interactions that enter, respectively,
in first and in second order perturbation theory.
(Here, additive means pairwise amongst the three atoms.)
In addition, we find that nonadditive dispersion interactions enter in second order
and that these contain a dependence on the geometrical configuration
of the three atoms.
(Nonadditive means that the terms appear collectively amongst the three atoms.)
The formalism is demonstrated by the calculation of the coefficients
for the  Li($2\,^2S$)-Li($2\,^2S$)-Li($2\,^{2}P$) system using
variationally generated atomic lithium wave functions in Hylleraas coordinates.
In addition, the  coefficients are given explicitly and as numerical values for the three basic geometrical configurations
of the nuclei in
an equilateral triangle, in an isosceles triangle, or equally spaced collinearly.
We show that the results are consistent with an available \textit{ab initio} quantum-chemical
calculation in the case of the equilateral triangle and equally spaced collinear configurations of the nuclei.

In general,
the long-range interactions~\cite{yan96, yan95, yan97} for three-body systems ~\cite{tang12, axilrod43, bell70, Marinescu97}
have many applications, such as in studies of atomic three-body recombination~\cite{soldan02, soldan03}, crystal structure~\cite{lotrich97}, color superfluids~\cite{rapp07}, Pfaffian states~\cite{paredes07}, and Efimov effects~\cite{gross09, pollack09}. Three-body recombination is a process in which three atoms collide and two of them form a molecule~\cite{soldan02, soldan03, daley09, lotrich97}. The binding energy among them depends not only on the additive but also on the nonadditive three-body dispersion forces~\cite{soldan02}. The nonadditive effects give a successful explanation of the failure of the Cauchy relations for the elastic constants in alkali halide crystals~\cite{bulski87}. Furthermore, the importance of these three-body interactions has been considered by many authors in connection with the Pfaffian state~\cite{paredes07} and the formation of trions in the color superfluidity~\cite{rapp07}.
And because of the large spatial extension and tiny binding energies, some properties of  Efimov states
would be expected to depend on the long-range  van der Waals interactions. Recently, some interesting theoretical investigations on the Efimov effect in higher partial waves ($P$-waves) have been reported~\cite{helfrich11}. However, additional research is warranted. In this paper, we will focus on the long-range three-body interactions for a system involving one atom in a $P$ state.

In order to emphasize the new aspect of the present analysis, in the remainder
of this section
we very briefly summarize some related work that places our analysis in context.
There are, of course, many studies on the ground and excited potential energy surfaces of three like alkali-metal atom
systems for spectroscopy and for chemical dynamics. We will refer to the case of
the homonuclear bound molecule as a trimer
and to the case of an atom and a bound homonuclear dimer (such as for
scattering) as an atom-dimer.
A  ``global" potential energy surface would encompass both of these regimes, but  a global potential
energy surface might not
be necessary for description of a particular physical process.
For example,
in a recent theoretical study of the photoassociation of a  $\mathrm{Cs} (6\,{}^2S)$ atom and a
$\mathrm{Cs}_2 (X{}^1\Sigma_g^+ ,v=0)$ dimer~\cite{PerRioLep15},
where two $\mathrm{Cs} (6\,{}^2S)$ atoms are already bound up as a diatomic molecule with vibrational quantum
number $v=0$,
long-range interaction potential energies ~\cite{LepDulKok10,LepVexBou11}
of the  (excited) atom-dimer system $\mathrm{Cs} (6\,{}^2P)$--$\mathrm{Cs}_2 (X{}^1\Sigma_g^+,v=0)$ were used.

Studies on three-lithium-atom system potential energy surfaces include
the trimer system~\cite{ThoIzmLem85,KraKeiSua99,MeyKeiKud01}
and treatments of the atom-dimer configuration within the context of a global trimer potential
energy surface~\cite{VarPai93,EhaYam99,GhaLarLar14}.
Motivated by ultra-cold science, some recent
studies on lithium focused on generation
of improved global potential energy surfaces for the trimer~\cite{brue0509,LiBruPar08,ByrMonMic09a}
or addressed the atom-dimer configuration in further detail~\cite{cvitas0501,ByrMonMic09b,ByrMicMon12},
with emphasis on the configuration of a $\mathrm{Li} (2\,{}^2S)$ atom
interacting with a (bound) lithium dimer for applications to scattering processes~\cite{cvitas0501,QueLauHon07,LiParBru08}.
Such investigations on long-range interactions for three-lithium-atom systems
continue to support  applications to atom-dimer scattering
at thermal energies~\cite{CopMatSte08}  and to atom-dimer photoassociation in the
ultra-cold energy domain~\cite{ByrMonMic09b}.
Calculations of  trimer excited electronic states are available
at various configurations and separations of the atoms~\cite{KraKeiSua99,EhaYam99,CviSolHut07,LiBruPar08,ByrMonMic09b,GhaLarLar14}.
Cvita\v{s} \textit{et al.}~\cite{cvitas06}
investigated  the case of three spin-polarized $\mathrm{Li} (2\,{}^2S)$ atoms
and considered the connection
between the long-range interactions of the atom-dimer and of the atom-atom-atom potential energy surfaces.
In contrast to their work, here we consider the atom-atom-atom case
where one of the lithium atoms is in the $2\,{}^2P$ excited
state.
As we will show,
our numerical results are consistent with the \textit{ab initio} quantum-chemical excited-state calculations of Ref.~\cite{CviSolHut07}
for the three atoms in the equilateral configuration  or in the equally-spaced collinear configuration
configuration.

\section{Theoretical Formulation}\label{The}

\subsection{The zeroth-order wave functions}\label{TheA}

The Hamiltonian for three well-separated (sufficiently far apart that electron exchange can be ignored) lithium atoms can be written as
\begin{equation}\label{e5}
{H}={H}^{(0)}+H' \,,
\end{equation}
where
\begin{eqnarray}
H^{(0)} &=& {H}_{1}^{(0)}+{H}_{2}^{(0)}+{H}_{3}^{(0)}\,, \\
H' & \equiv& V_{123} = V_{12}+V_{23}+V_{31}\,,
\end{eqnarray}
with $H_{1}^{(0)}$, $H_{2}^{(0)}$, and $H_{3}^{(0)}$, the unperturbed Hamiltonian of,
respectively, atom 1, 2, and 3 and
$V_{12}$, $V_{23}$, and $V_{31}$ their mutual electrostatic interactions.
We label the atoms by $I$, $J$, and $K$, with, respectively,
internal coordinates
$\boldsymbol{\sigma}$, $\boldsymbol{\rho}$, and $\boldsymbol{\varsigma}$.
When the labels $I$, $J$, or $K$ appear, it is understood that cyclic permutation
can be used.
$V_{IJ}$ can be expanded according to Refs.~\cite{bell70, fontana61}
\begin{equation}\label{e2}
V_{IJ}=\sum_{l_Il_J}\sum_{m_Im_J}T_\text{$l_I-m_I$}(\boldsymbol{\sigma})T_\text{$l_Jm_J$}(\boldsymbol{\rho})W_{l_Il_J}^{m_I-m_J}(IJ) \,.
\end{equation}
 In Eq.(\ref{e2}), the multipole transition operators are
\begin{eqnarray}
T_{l_I-m_I}(\boldsymbol{\sigma})&=&\sum_{i}Q_i\sigma ^{l_I}_{i}Y_{l_I-m_I}(\hat{\boldsymbol{\sigma}_i}) \,,  \\ \label{e3a}
T_{l_Jm_J}(\boldsymbol{\rho})&=&\sum_{j}q_j\rho ^{l_J}_{j}Y_{l_Jm_J}(\hat{\boldsymbol{\rho}_j}) \,,  \label{e3b}
\end{eqnarray}
and the geometry factor is
\begin{eqnarray}\label{e4}
W_{l_Il_J}^{m_I-m_J}(IJ)&=&\frac{4\pi(-1)^{l_J}}{R_{IJ}^{l_I+l_J+1}}\frac{(l_I+l_J-m_I+m_J)!(l_I,l_J)^{-1/2}}{[(l_I+m_I)!
(l_I-m_I)!(l_J+m_J)!(l_J-m_J)!]^{1/2}}
P_{l_I+l_J}^{m_I-m_J}(\cos\theta_{IJ})\nonumber\\
&\times & \exp[{i(m_I-m_J)\Phi_{IJ}}] \,,
\end{eqnarray}
where ${\bf R}_{IJ}={\bf R}_J-{\bf R}_I$ is the relative position vector from atom $I$ to atom $J$, the notation
$(a,b,\ldots)=(2a+1)(2b+1)\ldots$,
and $P_{l_I+l_J}^{m_I-m_J}(\cos\theta_{IJ})$ is the associated Legendre function with $\theta_{IJ}$ representing the angle between ${\bf R}_{IJ}$ and the $z$-axis. Similar expressions result for $V_{JK}$ and $V_{KI}$.
The choice of the $z$-axis is discussed below.

For the Li($n_0S$)-Li($n_0S$)-Li($n_0'L$) system where the angular momentum of one atom is $L$ and the associated magnetic quantum number is $M$, there are three orthogonal eigenvectors for the unperturbed Hamiltonian corresponding to the same energy eigenvalue $E_{n_0n_0n_0'}^{(0)}=2E_{n_0S}^{(0)}+E_{n_0'L}^{(0)}$,
\begin{eqnarray}
\ket{\phi_1}&=&\ket{\varphi_{n_0'}(LM;\boldsymbol{\sigma})\varphi_{n_0}(0;\boldsymbol{\rho})\varphi_{n_0}(0;\boldsymbol{\varsigma})}\,, \label{e6a_a} \\
\ket{\phi_2}&=&\ket{\varphi_{n_0}(0;\boldsymbol{\sigma})\varphi_{n_0'}(LM;\boldsymbol{\rho})\varphi_{n_0}(0;\boldsymbol{\varsigma})}\,, \label{e6b_b}
\\
\ket{\phi_3}&=&\ket{\varphi_{n_0}(0;\boldsymbol{\sigma})\varphi_{n_0}(0;\boldsymbol{\rho})\varphi_{n_0'}(LM;\boldsymbol{\varsigma})}\,.  \label{e6c_c}
\end{eqnarray}
According to degenerate perturbation theory, the zeroth-order wave function is a linear combination of the eigenvectors given in Eqs.~(\ref{e6a_a}), (\ref{e6b_b}), and (\ref{e6c_c}),
\begin{equation}\label{e7}
\ket{\Psi^{(0)}}=a\ket{\phi_1}+b\ket{\phi_2}+c\ket{\phi_3} \,.
\end{equation}
The expansion coefficients $a$, $b$, and $c$ are determined by diagonalizing the perturbation $V_{123}$ in the basis set $\{\phi_1, \phi_2, \phi_3\}$, which depends on the geometrical configuration formed by the three atoms.

\subsection{Choice of coordinates for three atoms}\label{TheB}

\begin{figure}
\begin{center}
\includegraphics[width=10cm,height=6cm]{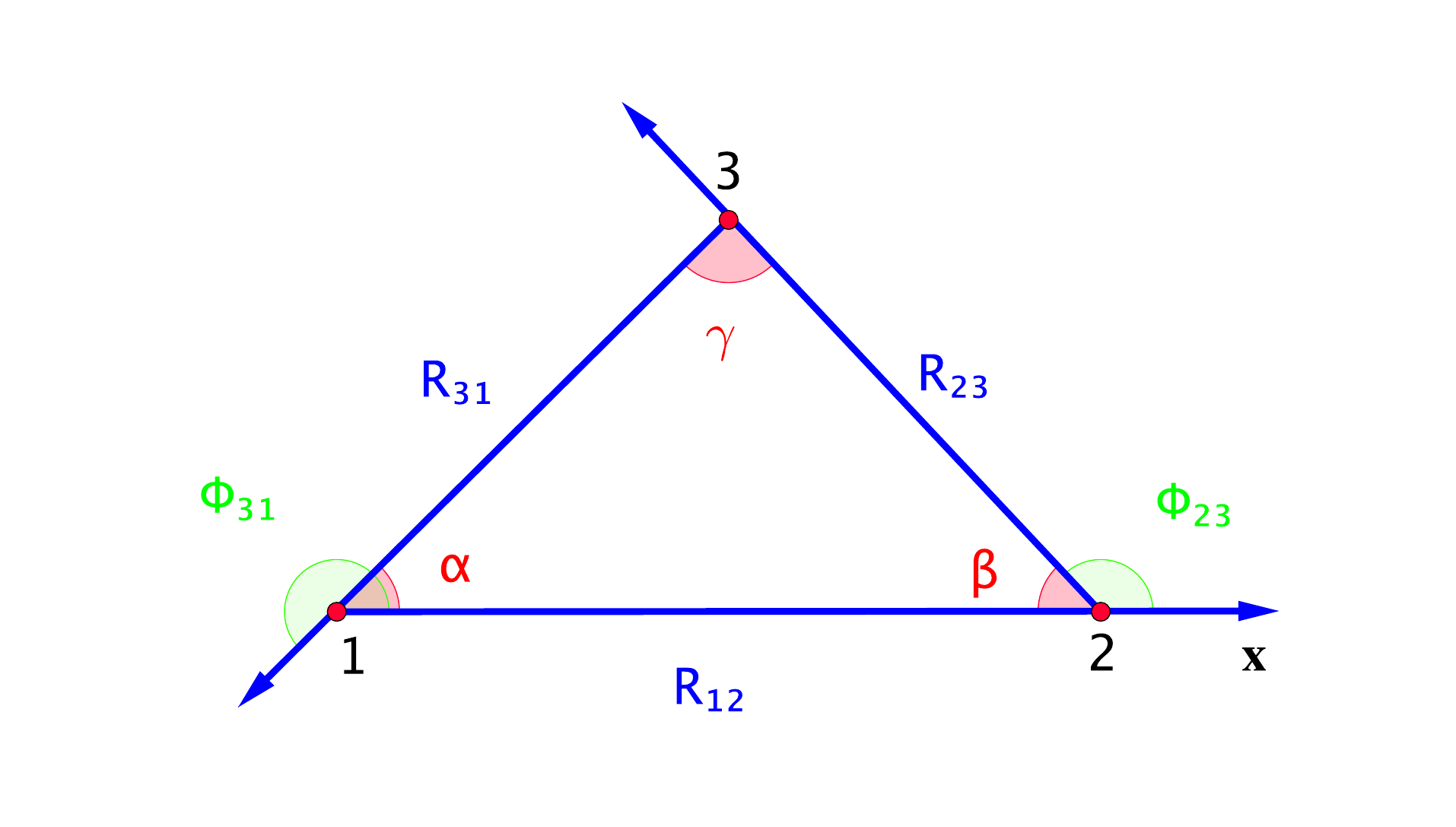}
\end{center}
\caption {\label{f1} Coordinate system for the three atoms: the \textit{z}-axis is perpendicular to the plane of the three nuclei and the \textit{x}-axis is parallel to ${\bf R}_{12}$. The angles satisfy $\Phi_{12}=0$, $\Phi_{23}=\pi-\beta$, $\Phi_{31}=\pi+\alpha$.
The nuclei lie in the $x$-$y$ plane.}
\label{figure_a1}
\end{figure}

In this work, we set the coordinates for the three atoms as shown in Fig.~\ref{f1}. Specifically, we choose the nucleus of atom~1 as the origin of our coordinate system and the $x$-$y$ plane as the plane formed by the three atomic nuclei. Furthermore, we set the \textit{x}-axis to be ${\bf R}_{12}$ and the \textit{z}-axis perpendicular to the \textit{x-y} plane by the right-hand convention. The interior angles of the triangle formed by the three atoms are denoted as $\alpha$, $\beta$, $\gamma$. Noting that~\cite{bell70} $\theta_{12}=\theta_{23}=\theta_{31}=\pi/2$, the associated Legendre functions can be simplified as:
\begin{eqnarray}
P_{l}^{m}(0)&=&\frac{1}{2^{l+1}}[1+(-1)^{l+m}](-1)^{\frac{l+m}{2}}(l+m)!
\bigg[\bigg(\frac{l+m}{2}\bigg)!\bigg]^{-1}\bigg[\bigg(\frac{l-m}{2}\bigg)!\bigg]^{-1} \,.
\end{eqnarray}
The angles $\Phi_{12}$, $\Phi_{23}$, and $\Phi_{31}$ satisfy $\Phi_{12}=0$, $\Phi_{23}=\pi-\beta$, and $\Phi_{31}=\pi+\alpha$, which can be used to simplify the exponential function $\exp[{i(m_I-m_J)\Phi_{IJ}}]$ in the geometry factor.

\subsection{The connection with other studies}\label{TheF}

With these eigenvectors and zero-order wave function, we can easily find the connection between this work and  preceding studies of long-range interactions. For example, if we eliminate the terms involving particle
3 (with angular momentum $L$), our formulas describe  the long-range interactions for the two-body $n_0S$-$n_0S$ system. If we set $a=\frac{1}{\sqrt{2}}$, $b=\pm\frac{1}{\sqrt{2}}$, $c=0$ and remove the terms involving particle 3 (with angular momentum 0), our  expressions describe  the long-range interactions for the two-body $n_0S$-$n_0'L$ system. If we set $a=1$, $b=c=0$ and $L=0$, the  long-range three-body $n_0S$-$n_0S$-$n_0'S$ interaction is described.

We noted earlier that Cvita\v{s} et al. \cite{cvitas06} considered the connection
between the long-range interactions for the atom-dimer and the atom-atom-atom potential energy surfaces
in the case  of the ground state for the three spin-aligned $\mathrm{Li} (2{}^2S)$ atoms.
They rewrote the distances separating the atoms in terms of Jacobi coordinates  $\tilde{r}$, $\tilde{R}$, and $\tilde{\theta}$,
where $\tilde{r}$ is the dimer internuclear distance, $\tilde{R}$ is the distance of the third atom from the center
of mass of the dimer, and $\tilde{\theta}$ is the angle between $\tilde{r}$ and $\tilde{R}$,
and expanded the corresponding results as power series in $\tilde{r}/\tilde{R}$ for $\tilde{r} \ll \tilde{R}$
obtaining
the contributions of the atom-atom coefficients to the atom-diatom coefficients.
The approximation is suitable because in the $n_0S$-$n_0S$-$n_0S$ system the atom-atom coefficients are constants.
However, for the $n_0S$-$n_0S$-$n_0^{\prime}L$ system, we find the atom-atom coefficients depend on the geometrical configuration of the three atoms.
In the present case, we can connect
the atom-dimer and trimer (atom-atom-atom) systems
by
adjusting the expansion coefficients in Eq.~(\ref{e7}).  Thus, if we set $a=b=0$, $c=1$, our formulas reduce to the long-range interactions between an atom $n_0'L$ and a diatom $n_0S$-$n_0S$.
However, unlike the study of Ref.~\cite{cvitas06}, (see also Ref.~\cite{VarPai93}), we do not include
short-range effects (damping), so in
our case the limit of an atom-dimer would still correspond
to the two atoms grouped as a dimer sufficiently separated that exchange is negligible.
In this work we
focus on the long-range interactions for the homonuclear atom-atom-atom lithium system.

In the Appendix, the general expressions for the $n_0S$-$n_0S$-$n_0'L$ cases
are presented up to second order in perturbation theory.
In the remainder of the main part of the paper we take $n_0=2=n_0'$, and $L=1$, to describe
the Li($2\,^2S$)-Li($2\,^2S$)-Li($2\,^{2}P$) system, for which we exhibit  and calculate
the coefficients.

\subsection{The first-order energy}\label{TheD}

According to perturbation theory, the first-order energy correction for the Li($2\,^2S$)-Li($2\,^2S$)-Li($2\,^2P$) system is
\begin{equation}\label{e14}
\Delta E^{(1)}=-\frac{C_3^{(12)}(1,M)}{R_{12}^3}-\frac{C_3^{(23)}(1,M)}{R_{23}^3}-\frac{C_3^{(31)}(1,M)}{R_{31}^3} \,,
\end{equation}
where
\begin{eqnarray}\label{e15}
C_{3}^{(IJ)}(1,M)&=& (A_{I}^{*}A_{J}+A_{J}^{*}A_{I}) \mathbb{D}_0(M) \,, \label{C3a}
\end{eqnarray}
\begin{eqnarray}\label{AI}
A_{1}=a, A_{2}=b, A_{3}=c\,,
\end{eqnarray}
and
\begin{eqnarray}\label{D0}
\mathbb{D}_0(M)&=&\frac{4\pi(-1)^{1+M}}{9(1-M)!(1+M)!}
|\langle\varphi_{n_0}(0;\boldsymbol{\sigma})\|T_1(\boldsymbol{\sigma})
\|\varphi_{n_0'}(1;\boldsymbol{\sigma})\rangle|^2\,.
\end{eqnarray}
In the above, $a$, $b$, $c$ are defined in Eq. (\ref{e7}). It should be mentioned that there exist only additive long-range
dipolar  interaction terms at this order of perturbation.

\subsection{The second-order energy}\label{TheE}

The second-order energy correction for the Li($2\,^2S$)-Li($2\,^2S$)-Li($2\,^2P$) system can be written as
\begin{eqnarray}\label{ee2}
\Delta E^{(2)}&=&-\sum_{n\geq 3}\bigg(\frac{C_{2n}^{(12)}(1,M)}{R_{12}^{2n}}+\frac{C_{2n}^{(23)}(1,M)}{R_{23}^{2n}}+\frac{C_{2n}^{(31)}(1,M)}{R_{31}^{2n}}
\nonumber \\
&+&\frac{C_{2n}^{(12,23)}(1,M)}
{R_{12}^nR_{23}^n}+\frac{C_{2n}^{(23,31)}(1,M)}{R_{23}^nR_{31}^n}+\frac{C_{2n}^{(31,12)}(1,M)}{R_{31}^nR_{12}^n}\bigg) \,,
\end{eqnarray}
where the $C_{2n}^{(12)}(1,M)$, $C_{2n}^{(23)}(1,M)$, and $C_{2n}^{(31)}(1,M)$ are the additive dispersion coefficients, and $C_{2n}^{(12,23)}(1,M)$, $C_{2n}^{(23,31)}(1,M)$, and $C_{2n}^{(31,12)}(1,M)$ are the nonadditive dispersion coefficients. The derivation of these coefficients is given in the Appendix. In this work we are only concerned with $n=3$ and 4 in Eq.~(\ref{ee2}). The corresponding dispersion coefficients are
\begin{eqnarray}
C_{6}^{(IJ)}(1,M)&=&(|A_{I}|^2 + |A_{J}|^2)\mathbb{D}_1(M) +|A_{K}|^2\mathbb{D}_2  \,, \label{C6a}
\end{eqnarray}
\begin{eqnarray}
C_{8}^{(IJ)}(1,M)&=&(|A_{I}|^2+|A_{J}|^2)\mathbb{Q}_1(M)+|A_{K}|^2\mathbb{Q}_2+(A_{I}^{*}A_{J}+ A_{J}^{*}A_{I})\mathbb{Q}_3(M) \,, \label{C8a}
\end{eqnarray}
\begin{eqnarray}
C_{6}^{(IJ,JK)}(1,M)&=&\mathbb{Q}_4(A_{K},A_{I},1,M,\eta_{J})  \,, \label{C6d}
\end{eqnarray}
\begin{eqnarray}
C_{8}^{(IJ,JK)}(1,M)&=&\mathbb{Q}_4(A_{K},A_{I},2,M,\eta_{J}) \,, \label{C8d}
\end{eqnarray}
with
\begin{eqnarray}\label{eta}
A_{1}=a, A_{2}=b, A_{3}=c, \eta_{1}=\alpha, \eta_{2}=\beta, \eta_{3}=\gamma\,,
\end{eqnarray}
where $a$, $b$ and $c$ are defined in Eq. (\ref{e7}), $\alpha$, $\beta$ and $\gamma$ are the interior angles,
and the other terms in Eqs. (\ref{C6a})-(\ref{C8d}) are given by
\begin{eqnarray}
\mathbb{D}_1(M)&=&\sum_{n_sn_t L_s}F_1(n_s,n_t,L_s,1;1,1;1,M) \,, \\
\mathbb{D}_2&=&\sum_{n_sn_t} F_2(n_s,n_t,1,1) \,,\\
\mathbb{Q}_1(M)&=& \sum_{n_sn_t L_s} [F_1(n_s,n_t,L_s,1;1,3;1,M) + F_1(n_s,n_t,L_s,1;2,2;1,M) \nonumber \\
&+&F_1(n_s,n_t,L_s,1;3,1;1,M) + F_1(n_s,n_t,L_s,2;1,1;1,M)] \,, \\
\mathbb{Q}_2&=&\sum_{n_sn_t}[F_2(n_s,n_t,1,2)+F_2(n_s,n_t,2,1)]  \,, \\
\mathbb{Q}_3(M)&=& \sum_{n_sn_t}[F_3(n_s,n_t,1,1;2,2;1,M)+F_3(n_s,n_t,1,2;2,1;1,M) \nonumber \\
&+&F_3(n_s,n_t,2,1;1,2;1,M)+F_3(n_s,n_t,2,2;1,1;1,M)] \,,
\end{eqnarray}
and
\begin{eqnarray}
\mathbb{Q}_4(A_K,A_I,\lambda,M,\eta_J)&=&2\sum_{n_tM_t}\operatorname{Re}[
A_K^*A_I\,e^{i(M_{t}-M)\eta_J}]F_4(n_t,\lambda,M_t;1,M) \,.\label{QQ4}
\end{eqnarray}
In the above $(I,J,K)$ forms a permutation of (1,2,3)
and the $F_i$-functions are defined by Eqs.~(\ref{AF1}), (\ref{AF2}), (\ref{AF3}), and (\ref{AF4}) in the Appendix.

\subsection{Three special geometrical configurations}\label{TheC}

In this work, three special geometrical configurations for the three atoms are considered. The first configuration is the equilateral triangle, where the interatomic separations are the same: $R_{12}=R_{23}=R_{31}=R$.
With this configuration, all the diagonal perturbation matrix elements are zero and all the off diagonal matrix elements are the same. The perturbation matrix with respect to $\{\phi_1,\phi_2,\phi_3\}$ thus becomes
\begin{equation}\label{e10}
{H}^{\prime}=H_{12}^{\prime}
\left(
  \begin{array}{ccc}
    0 & 1 & 1\\
    1 & 0 & 1\\
    1 & 1 & 0\\
  \end{array}
\right)\,,
\end{equation}
where
\begin{eqnarray}
H^{\prime}_{12}
  &=&\frac{4\pi}{R^{2L+1}}\frac{(-1)^{M}[(2L-1)!!]^2}{(2L+1)^2(L-M)!(L+M)!} |\langle\varphi_{n_0}(0;\boldsymbol{\sigma})\|T_L(\boldsymbol{\sigma})
\|\varphi_{n_0'}(L;\boldsymbol{\sigma})\rangle|^2\,.
\end{eqnarray}
Solving the eigenvalue problem of the above matrix, one obtains the eigenvalues: $2H_{12}^{\prime}$, $-H_{12}^{\prime}$, $-H_{12}^{\prime}$, and the corresponding orthonormalized zeroth-order wave functions:
\begin{eqnarray}
\Psi_{1,\Delta}^{(0)}&=&\frac{1}{\sqrt{3}}|\phi_1\rangle+\frac{1}{\sqrt{3}}|\phi_2\rangle+\frac{1}{\sqrt{3}}|\phi_3\rangle \,, \label{delta1} \\
\Psi_{2,\Delta}^{(0)}&=&\frac{1}{\sqrt{2}}|\phi_1\rangle-\frac{1}{\sqrt{2}}|\phi_3\rangle \,, \label{delta2} \\
\Psi_{3,\Delta}^{(0)}&=&\frac{1}{\sqrt{6}}|\phi_1\rangle-\sqrt{\frac{2}{3}}|\phi_2\rangle+\frac{1}{\sqrt{6}}|\phi_3\rangle \,, \label{delta3}
\end{eqnarray}
where the symbol $\Delta$ denotes the equilateral triangle.

The second special geometrical configuration is the isosceles right triangle such that
$R_{12}=\frac{1}{\sqrt{2}}R_{23}=R_{31}=R$.
With this configuration, the perturbation matrix has the form
\begin{eqnarray}\label{e11}
{H}^{\prime}&=&H_{12}^{\prime}
\left(
  \begin{array}{ccc}
    0 & 1 & 1\\
    1 & 0 & \frac{1}{2\sqrt{2}}\\
    1 & \frac{1}{2\sqrt{2}} & 0\\
  \end{array}
\right)\,.
\end{eqnarray}
The eigenvalues are $\frac{H_{12}^{\prime}}{8} \left(\sqrt{2}+\sqrt{130}\right), \frac{H_{12}^{\prime}}{8} \left(\sqrt{2}-\sqrt{130}\right), -\frac{H_{12}^{\prime}}{2 \sqrt{2}}$, and the corresponding orthonormalized zeroth-order wave functions are
\begin{eqnarray}
\Psi_{1,\bot}^{(0)}&=&\frac{\sqrt{130}-\sqrt{2}}{2 \sqrt{65-\sqrt{65}}}|\phi_1\rangle+\frac{4}{\sqrt{65-\sqrt{65}}}|\phi_2\rangle
+\frac{4}{\sqrt{65-\sqrt{65}}}|\phi_3\rangle  \,,  \label{bot1} \\
\Psi_{2,\bot}^{(0)}&=&\frac{-(\sqrt{130}+\sqrt{2})}{2 \sqrt{65+\sqrt{65}}}|\phi_1\rangle+\frac{4}{\sqrt{65+\sqrt{65}}}|\phi_2\rangle
+ \frac{4}{\sqrt{65+\sqrt{65}}}|\phi_3\rangle \,,  \label{bot2} \\
\Psi_{3,\bot}^{(0)}&=&-\frac{1}{\sqrt{2}}|\phi_2\rangle+\frac{1}{\sqrt{2}}|\phi_3\rangle \,,  \label{bot3}
\end{eqnarray}
where the symbol $\bot$ denotes the isosceles right triangle.

The third special geometrical configuration is that the three atoms are in a straight line
such that $R_{12}=\frac{1}{2}R_{23}=R_{31}=R$.
With this configuration, the perturbation matrix is
\begin{equation}\label{e12}
{H}^{\prime}=H_{12}^{\prime}
\left(
  \begin{array}{ccc}
    0 & 1 & 1\\
    1 & 0 & \frac{1}{8}\\
    1 & \frac{1}{8} & 0\\
  \end{array}
\right)\,.
\end{equation}
The eigenvalues are $\frac{H_{12}^{\prime}}{16} \left(1+3 \sqrt{57}\right), \frac{H_{12}^{\prime}}{16} \left(1-3 \sqrt{57}\right), -\frac{H_{12}^{\prime}}{8}$, and the corresponding orthonormalized zeroth-order wave functions are
\begin{eqnarray}
\Psi_{1,\text{---}}^{(0)}&=&\frac{3\sqrt{57}-1}{\sqrt{1026-6\sqrt{57}}}|\phi_1\rangle+\frac{16}{\sqrt{1026-6\sqrt{57}}}|\phi_2\rangle
+\frac{16}{\sqrt{1026-6\sqrt{57}}}|\phi_3\rangle  \,,  \label{line1}\\
\Psi_{2,\text{---}}^{(0)}&=&\frac{-(3\sqrt{57}+1)}{\sqrt{1026+6\sqrt{57}}}|\phi_1\rangle+\frac{16}{\sqrt{1026+6\sqrt{57}}}|\phi_2\rangle
+\frac{16}{\sqrt{1026+6\sqrt{57}}}|\phi_3\rangle  \,,  \label{line2} \\
\Psi_{3,\text{---}}^{(0)}&=&-\frac{1}{\sqrt{2}}|\phi_2\rangle + \frac{1}{\sqrt{2}}|\phi_3\rangle \,,  \label{line3}
\end{eqnarray}
where the symbol $\text{--}$ denotes the geometrical configuration of a straight line.

\section{Results and Discussion}\label{Res}

In the present work, the atomic wave functions of lithium were constructed variationally using Hylleraas basis sets
and the intermediate states were generated by diagonalizing the lithium Hamiltonian, as in Ref.~\cite{tang09}. All relevant
matrix elements of the multipole transition operators were thus calculated, including the finite nuclear mass corrections. With these,
we calculate
the first-order dipolar and second-order long-range dispersion coefficients for
the Li${(2\,^2S)}$-Li${(2\,^2S)}$-Li${(2\,^2P)}$ system.

Table~\ref{TabI} lists the values of $\mathbb{D}_0(M=0)$, $\mathbb{D}_0(M=\pm 1)$, $\mathbb{D}_1(M=0)$, $\mathbb{D}_1(M=\pm1)$, $\mathbb{D}_2$, $\mathbb{Q}_1(M=0)$, $\mathbb{Q}_1(M=\pm1)$, $\mathbb{Q}_2$, $\mathbb{Q}_3(M=0)$, and $\mathbb{Q}_3(M=\pm 1)$ for lithium isotopes, all of which are
independent of the geometrical configuration of the three lithium atoms. However, the quantity $\mathbb{Q}_4$, not listed in Table~\ref{TabI}, is related to the nonadditive dispersion coefficients and is  dependent on the geometrical configuration.
$\mathbb{D}_0(M=0)$ and $\mathbb{D}_0(M=\pm 1)$ are connected with the first-order additive coefficient $C_3^{(IJ)}(L,M)$. $\mathbb{D}_1(M=0)$, $\mathbb{D}_1(M=\pm1)$, and $\mathbb{D}_2$ are connected with the second-order additive dispersion coefficient $C_6^{(IJ)}(L,M)$, and $\mathbb{Q}_1(M=0)$, $\mathbb{Q}_1(M=\pm1)$, $\mathbb{Q}_2$, $\mathbb{Q}_3(M=0)$, and $\mathbb{Q}_3(M=\pm1)$ are connected with the second-order additive dispersion coefficient $C_8^{(IJ)}(L,M)$.

With the values in Table~\ref{TabI}, we can obtain the long-range interaction coefficients for geometrical configurations
specified by $R_{12}$, $R_{23}$, $R_{31}$, $\alpha$, $\beta$, and $\gamma$ as follows:
Initially, we obtain the geometric parameters for the configuration under consideration by the method as described in
Sec.~\ref{TheC}. Then, the long-range interaction coefficients for that configuration are given by the Eqs.~(\ref{e14})-(\ref{QQ4}). In the following, we will discuss the long-range coefficients by explicitly
evaluating coefficients for the three elementary geometrical configurations
of the atoms in an equilateral triangle, an  isosceles triangle, or equally-spaced collinearly, representing,
respectively, the symmetries $D_{3h}$, $C_{2v}$, and $D_{\infty h}$.

\subsection{Dipolar and dispersion coefficients for an equilateral triangle}\label{res1}

Using the coefficients ${a,b,c}$ of the zeroth-order wave functions Eqs.~(\ref{delta1})-(\ref{delta3}), for the case where
the three atoms form an equilateral triangle, the first-order dipolar coefficients are listed in Table~\ref{TabII} and the second-order additive and nonadditive dispersion coefficients are listed in Tables~\ref{TabIII}-\ref{TabVI}.

From Table~\ref{TabII}, we can see that for the zeroth-order wave function $\Psi_{1,\Delta}^{(0)}$, $C_3^{(12)}(1,M=0)$, $C_3^{(23)}(1,M=0)$, and $C_3^{(31)}(1,M=0)$ are all the same because  $a=b=c=1/\sqrt{3}$; similarly $C_3^{(12)}(1,M=\pm 1)$, $C_3^{(23)}(1,M=\pm 1)$, and
$C_3^{(31)}(1,M=\pm 1)$ are all the same. For the zeroth-order wave function $\Psi_{2,\Delta}^{(0)}$, $C_3^{(12)}(1,M)$ and $C_3^{(23)}(1,M)$, whenever $M=0$ or $M=\pm 1$, are zero because $b=0$. For the zeroth-order wave function $\Psi_{3,\Delta}^{(0)}$, $C_3^{(12)}(1,M)$ and $C_3^{(23)}(1,M)$ are the same because $a=c=1/\sqrt{6}$. The coefficients between $M=0$ and $M=\pm 1$ satisfy the relationship $C_3^{(IJ)}(1,M=0)=$$-$$2C_3^{(IJ)}(1,M=\pm 1)$. Also listed in Table~\ref{TabII} are the values for $^7$Li and $^6$Li obtained by taking the finite
nuclear mass into consideration.

For the leading terms of the second-order long-range interaction, there exist both additive and nonadditive terms
as shown in Table~\ref{TabIII} for $C_6^{(IJ)}(1,M=0)$ and $C^{(IJ,JK)}_6(1,M=0)$ and in Table~\ref{TabIV} for $C_6^{(IJ)}(1,M=\pm 1)$
and $C_6^{(IJ,JK)}(1,M=\pm 1)$, for the Li($2\,^2S$)-Li($2\,^2S$)-Li($2\,^{2}P$) system.
For the second zeroth-order wave function $\Psi_{2,\Delta}^{(0)}$, the nonadditive coefficients $C_6^{(23,31)}(1,M)$ and $C_6^{(31,12)}(1,M)$ are zero for $M=0$ or $M=\pm 1$ because  $\mathbb{Q}_4(a,b,1,M,\gamma)=0$ and $b=0$. For $\Psi_{2,\Delta}^{(0)}$ and $\Psi_{3,\Delta}^{(0)}$, we have $C_6^{(12)}(1,M)=C_6^{(23)}(1,M)$ and $C_6^{(23,31)}(1,M)=C_6^{(31,12)}(1,M)$ for $M=0$ or $M=\pm 1$  because $a=c$.

The long-range dispersion coefficients $C_8^{(IJ)}(1,M)$ and $C_8^{(IJ,JK)}(1,M)$ are listed in Table~\ref{TabV} for $M=0$ and in Table ~\ref{TabVI} for $M=\pm 1$ for the Li($2\,^2S$)-Li($2\,^2S$)-Li($2\,^{2}P$) system. These coefficients have very similar characteristic
as $C_6^{(IJ)}(1,M)$ and $C_6^{(IJ,JK)}(1,M)$.
For example, for fixed $M$, $C_8^{(IJ)}(1,M)$ are all the same for the zeroth-order wave function $\Psi_{1,\Delta}^{(0)}$; similarly for $C_8^{(IJ,JK)}(1,M)$, they are all the same for $\Psi_{1,\Delta}^{(0)}$. For $\Psi_{2,\Delta}^{(0)}$, $C_8^{(23,31)}(1,M)$ and  $C_8^{(31,12)}(1,M)$ are zero because $\mathbb{Q}_4(a,b,2,M,\gamma)=0$ and $b=0$. We also have $C_8^{(12)}(1,M)=C_8^{(23)}(1,M)$ and $C_8^{(23,31)}(1,M)=C_8^{(31,12)}(1,M)$ for both $\Psi_{2,\Delta}^{(0)}$ and $\Psi_{3,\Delta}^{(0)}$.

\begin{figure}
\begin{center}
\includegraphics[width=14cm,height=8cm]{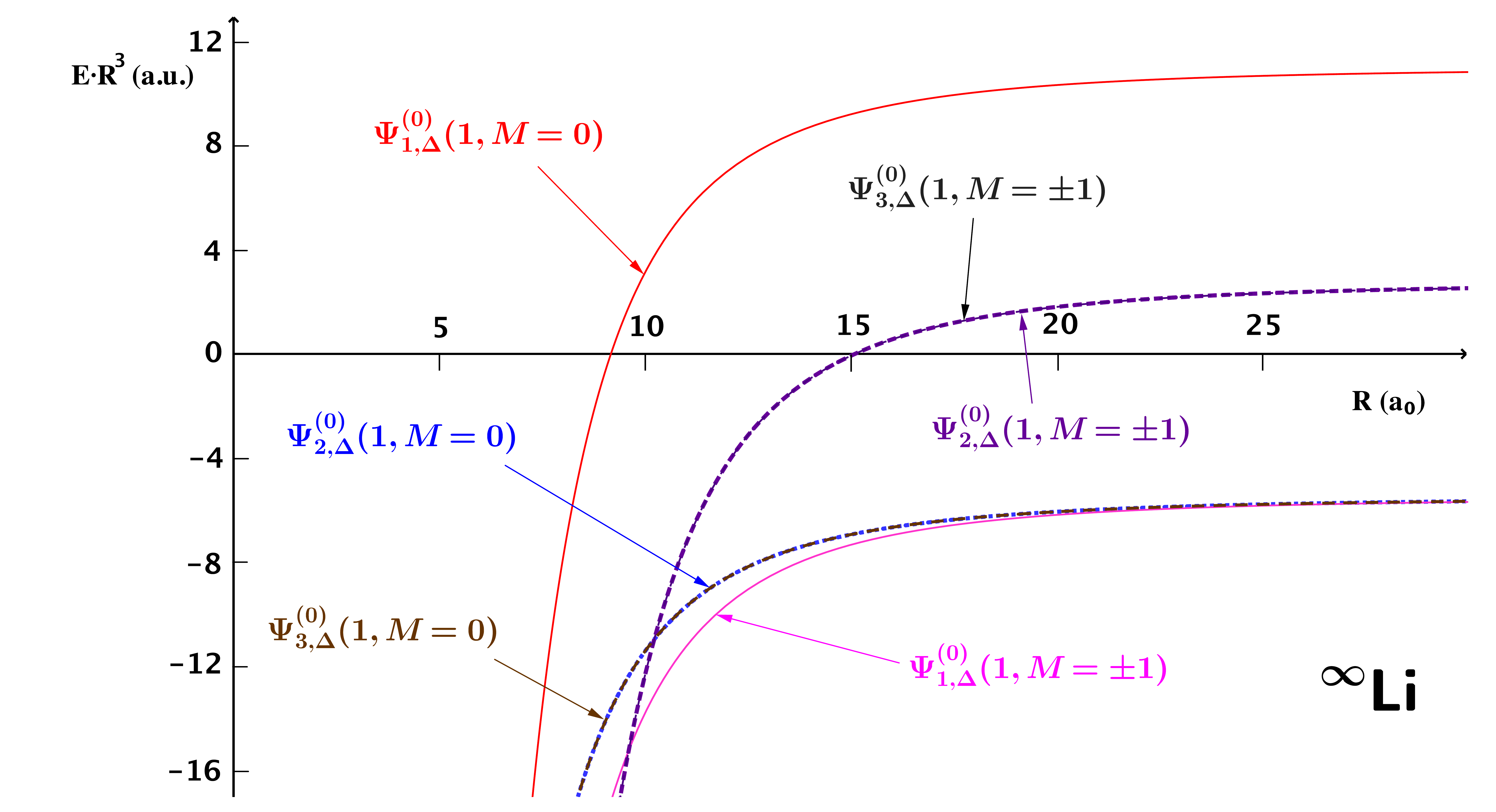}
\end{center}
\caption {\label{f2} Long-range potentials $(E)$ multiplied by $R^3$ for the Li($2\,^2S$)-Li($2\,^2S$)-Li($2\,^{2}P$) system for three different types of the zeroth-order wave functions, where the three atoms form an equilateral triangle and $R$ is the interatomic distance, in atomic units.
For each curve labeled by a wave function, the plotted curve is the sum of $\Delta E^{(1)}$ and $\Delta E^{(2)}$.
}
\end{figure}

From Tables~\ref{TabIII}-\ref{TabVI}, we can also see that the dispersion coefficients for the additive terms are always positive,
but the dispersion coefficients for the nonadditive terms can be positive or negative. Furthermore, the absolute values of the non-zero nonadditive dispersion coefficients are less than the additive dispersion coefficients by one to two orders of  magnitude. However, the nonadditive terms may not be neglected in constructing an accurate potential surface for Li($2\,^2S$)-Li($2\,^2S$)-Li($2\,^{2}P$). For example, for the case of $\Psi_{2,\Delta}^{(0)}$, the ratio of $(\frac{C_8^{(12,23)}(1,M=\pm 1)}{R_{12}^4R_{23}^4})/(\frac{C_8^{(12)}(1,M=\pm 1)}{R_{12}^8})
\simeq 18\%$.
For this configuration the energies $\Delta E^{(1)}$ and $\Delta E^{(2)}$  are given by, respectively,
Eqs.~(\ref{e14}) and (\ref{ee2}), and are listed in Table~\ref{TabII-add}.
The curves of potential energy $(E)$ multiplied by $R^3$ corresponding to the different zeroth-order wave functions are plotted in Figure~\ref{f2}.

We identify the $\Psi_{1,\Delta}^{(0)}$, $\Psi_{2,\Delta}^{(0)}$, and $\Psi_{3,\Delta}^{(0)}$ states
with $M=0$
as trimer states of $A$ symmetry and those with $M=\pm1$ as trimer states of $E$ symmetry
by comparison with the \textit{ab initio} calculations of Ref.~\cite{CviSolHut07},
where $A$ and $E$ are standard nomenclature labeling $D_{3h}$ symmetry in electronic states.
This identification is consistent with the double-degeneracy of $E$ symmetry states.
In Ref.~\cite{CviSolHut07}, quantum-chemical calculations were carried out for the ground and excited quartet electronic states of
$\mathrm{Li}_3$. While the emphasis of that work was on the ground electronic state,
a plot (Fig.~2 of Ref.~\cite{CviSolHut07}) is given of the excited state potential energies
corresponding to Li($2\,^2S$)-Li($2\,^2S$)-Li($2\,^{2}P$)
for the atoms
in an equilateral configuration at atomic separations up to $10^{-9}~\mathrm{m}$ $(19~a_0)$.
Six electronic states are given, of which three are identified with $A$ symmetry and three with $E$ symmetry.
At $R\approx 19~a_0$, $\Delta E^{(1)}$ is the leading contribution
to the long-range interactions and
we note that the ordering
of the states from lowest to highest energy at the maximum distance is in accord
with our results shown in Table~\ref{TabII-add}.
At this value of $R$, our values of $\Delta E^{(1)}$ range from $-350~\mathrm{cm}^{-1}$
to $175~\mathrm{cm}^{-1}$ and are entirely consistent with the results
shown in Fig.~2 of Ref.~\cite{CviSolHut07}.
Because the calculation of Ref.~\cite{CviSolHut07} includes exchange
a more quantitative
comparison may be inconclusive.
In addition, because our choice of coordinate system  (see Fig.~\ref{f1}) doesn't naturally
reflect the symmetries of the equilateral triangle, we refrain from making further symmetry
assignments.

\subsection{Dipolar and dispersion coefficients for an isosceles right triangle}\label{res2}

For the configuration of isosceles right triangle, the first-order dipolar coefficients are listed in Table~\ref{TabVII} and the second-order additive and nonadditive dispersion coefficients are listed in Tables~\ref{TabVIII}-\ref{TabXI}.

For this configuration, we have $b=c$ for the first two zeroth-order wave functions $\Psi_{1,\bot}^{(0)}$ and $\Psi_{2,\bot}^{(0)}$, and $b=-c$ and $a=0$ for the third zeroth-order wave function $\Psi_{3,\bot}^{(0)}$. We can clearly see that $C_3^{(12)}(1,M)$ equals $C_3^{(31)}(1,M)$ for $\Psi_{1,\bot}^{(0)}$ and $\Psi_{2,\bot}^{(0)}$, and $C_3^{(12)}(1,M)=C_3^{(31)}(1,M)=0$ for $\Psi_{3,\bot}^{(0)}$, as shown in Table~\ref{TabVII}. Compared to the values in Table~\ref{TabII}, we can conclude that a change in geometric configuration will influence the long-range coefficients.

\begin{figure}
\begin{center}
\includegraphics[width=14cm,height=8cm]{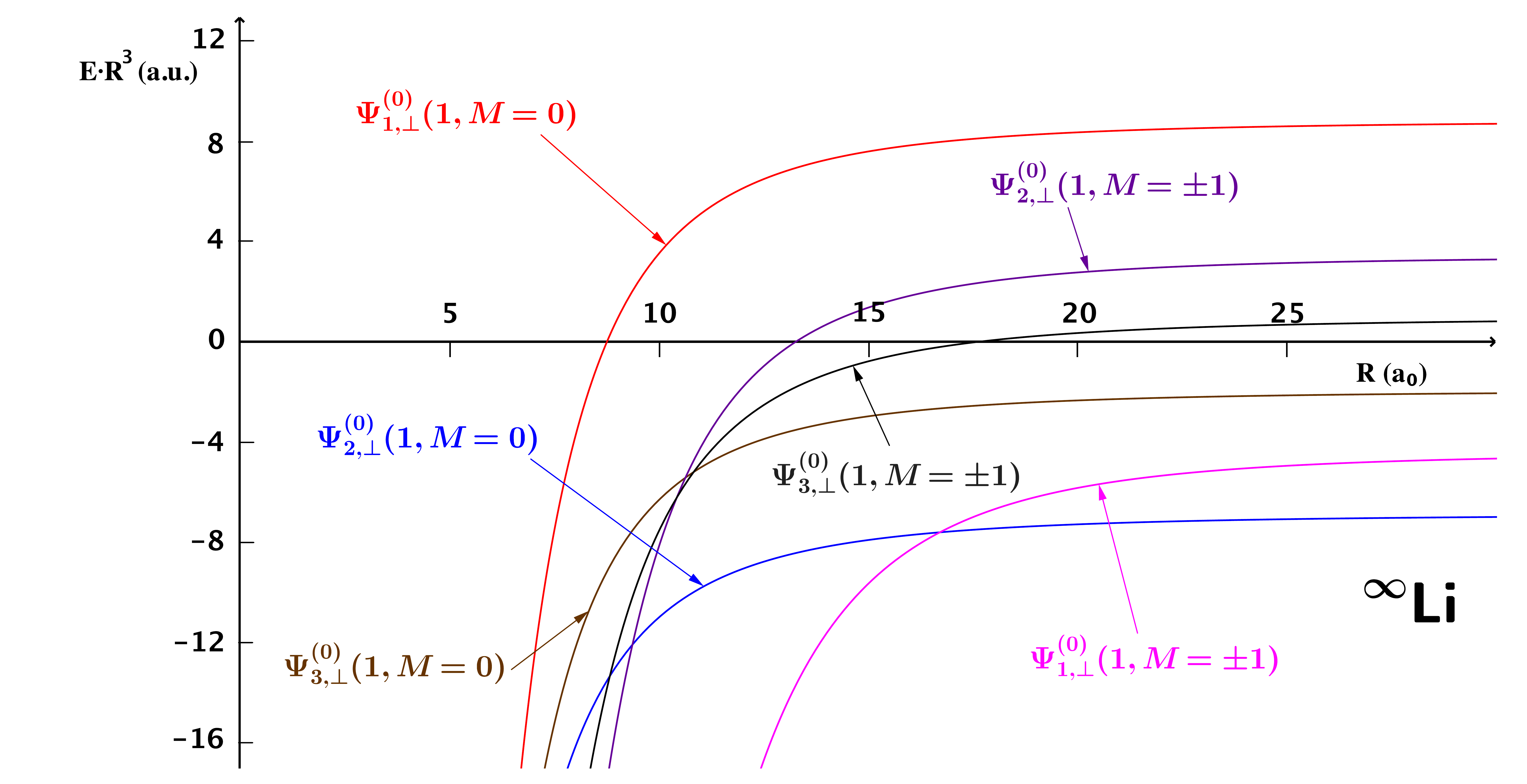}
\end{center}
\caption {\label{f3} Long-range potentials $(E)$ multiplied by $R^3$ for the Li($2\,^2S$)-Li($2\,^2S$)-Li($2\,^{2}P$) system for three different types of the zeroth-order wave functions, where the three atoms form an isosceles right triangle and $R$ is the interatomic distance on the congruent sides, in atomic units.
For each curve labeled by a wave function, the plotted curve is the sum of $\Delta E^{(1)}$ and $\Delta E^{(2)}$.
}
\end{figure}

The second-order dispersion coefficients $C_6^{(IJ)}$, $C_6^{(IJ,JK)}$, $C_8^{(IJ)}$, and $C_8^{(IJ,JK)}$ are listed in Tables~\ref{TabVIII}-\ref{TabXI}. One can see that $C_6^{(12)}(1,M)=C_6^{(31)}(1,M)$, $C_6^{(12,23)}(1,M)=C_6^{(23,31)}(1,M)$, $C_8^{(12)}(1,M)=C_8^{(31)}(1,M)$, and $C_8^{(12,23)}(1,M)=C_8^{(23,31)}(1,M)$ because $b=\pm c$, $a=0$, and $\beta=\gamma$. Also $C_8^{(31,12)}(1,M)=0$ and $C_6^{(31,12)}(1,M)\neq 0$ because $\alpha=\pi/2$ and  $M_t-M$
can be even or odd in Eq.~(\ref{QQ4}).
We find
that allowing for finite nuclear mass increases the additive dispersion coefficients,
as shown in Tables~\ref{TabVIII}-\ref{TabXI}.
Similarly to Sec.~\ref{res1}, the nonadditive terms may not be neglected in constructing a three-body potential surface for Li($2\,^2S$)-Li($2\,^2S$)-Li($2\,^{2}P$). The curves of potential energy $(E)$ multiplied by $R^3$ resulting
from $\Delta E^{(1)}$ and $\Delta E^{(2)}$ for this geometrical structure are shown in Fig.~\ref{f3}.

\subsection{Dipolar and dispersion coefficients for a straight line}\label{res3}

\begin{figure}
\begin{center}
\includegraphics[width=14cm,height=8cm]{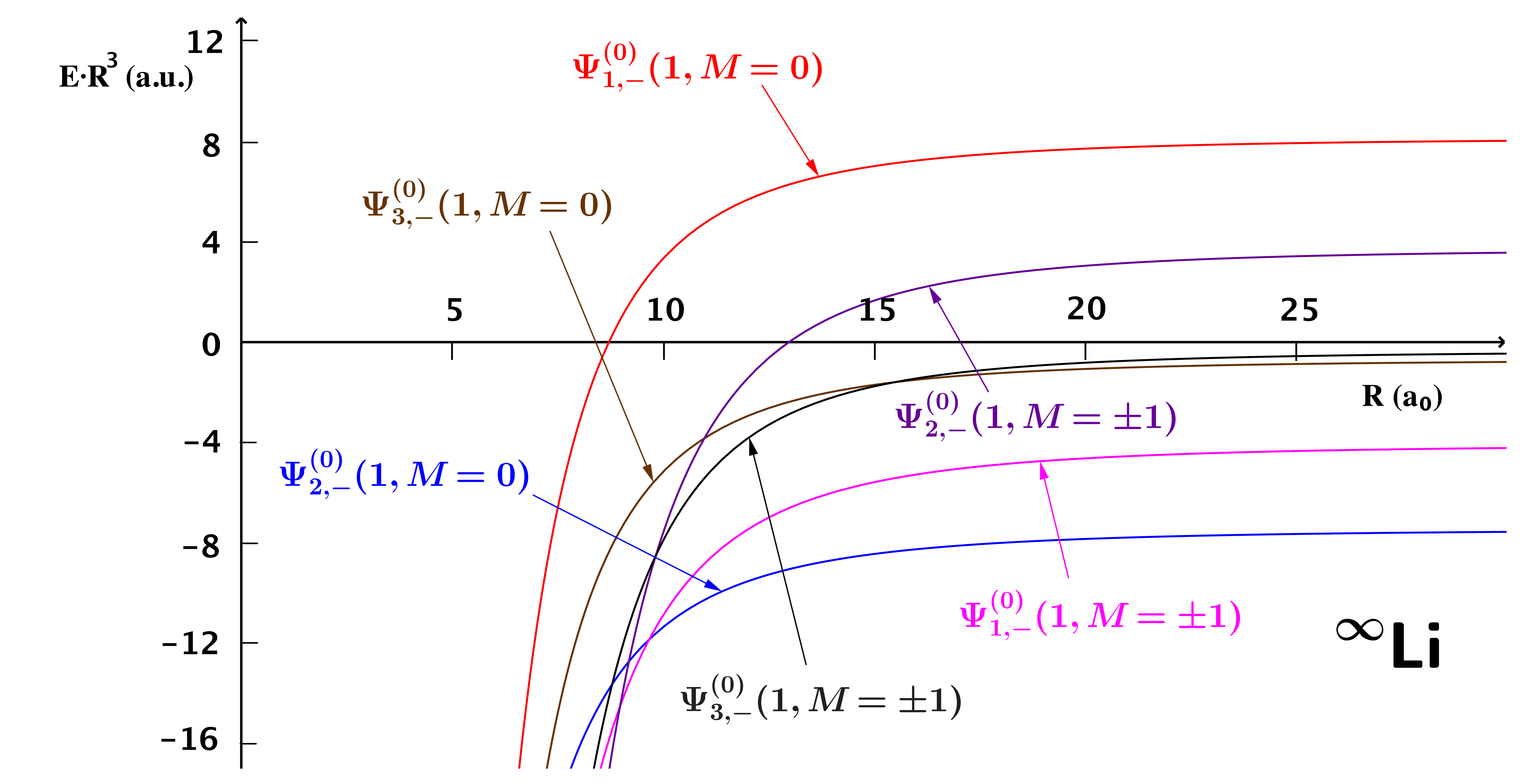}
\end{center}
\caption {\label{f4} Long-range potentials $(E)$ multiplied by $R^3$ for the Li($2\,^2S$)-Li($2\,^2S$)-Li($2\,^{2}P$) system for three different types of the zeroth-order wave functions, where the three atoms form a straight line and $R$ is the distance between adjacent atoms, in atomic units.
For each curve labeled by a wave function, the plotted curve is the sum of $\Delta E^{(1)}$ and $\Delta E^{(2)}$.
}
\end{figure}

For the configuration of three atoms equally spaced
and forming a straight line, the long-range dipolar and dispersion coefficients are listed in Tables~\ref{TabXII}-\ref{TabXVI}.
Since the zeroth-order wave function coefficients have $b=c$ in Eqs.~(\ref{line1}) and (\ref{line2}), and $a=0$ and $b=-c$ in Eq.~(\ref{line3}) the dispersion coefficients have similar characteristics as the case of the isosceles right triangle of Sec.~\ref{res2}.
The only differences are the values of three interior angles: $\beta=\gamma=0$ and $\alpha=2\pi$, which leads to the relatively larger nonzero dispersion
coefficients  $C_8^{(31,12)}(1,M=0)$ and $C_8^{(31,12)}(1,M=\pm1)$.
The corresponding curves of  potential energy $(E)$ multiplied by $R^3$ resulting from  $\Delta E^{(1)}$ and $\Delta E^{(2)}$  are shown in Fig.~\ref{f4}.

As in Sec.~\ref{res1}, we can make a comparison with the results of Ref.~\cite{CviSolHut07},
where, in their Fig.~1, the excited electronic potential energies of Li($2\,^2S$)-Li($2\,^2S$)-Li($2\,^{2}P$)
are given for the equally spaced collinear geometry at values of $R$ up to $19~a_0$.
We identify $\Psi_{1,\text{--}}^{(0)}$, $\Psi_{2,\text{--}}^{(0)}$, and $\Psi_{3,\text{--}}^{(0)}$ states
with $M=0$
as trimer states of $\Sigma$ symmetry and those with $M=\pm1$ as trimer states of $\Pi$ symmetry,
corresponding to $D_{\infty h}$ symmetry labels.
In the present work, the magnitude of the leading long-range interaction energy  $\Delta E^{(1)}$
ranges from $-341~\mathrm{cm}^{-1}$ to $170~\mathrm{cm}^{-1}$
and is consistent with the \textit{ab initio} results shown in Fig.~1 of Ref.~\cite{CviSolHut07}.

\section{Conclusion}\label{Con}

The long-range additive dipolar and additive dispersion interactions and nonadditive  dispersion interactions for the Li($2\,^2S$)-Li($2\,^2S$)-Li($2\,^{2}P$) system were obtained using perturbation theory.
The additive dipolar and additive dispersion interactions and nonadditive dispersion interactions depend on the geometrical configuration of the atoms.

Here we found that the nonadditive dispersion interactions start to appear at the \textit{second order} in the perturbative
treatment, which is different from the case of three $S$ atoms where the geometry dependent nonadditive dispersion interactions start to appear at the \textit{third order}. While the formulas apply to all geometrical configurations,
we demonstrated the methodology for three basic types of geometrical configurations
(nuclei forming an equilateral triangle or an isosceles triangle, or nuclei equally-spaced and collinear)
by calculating coefficients to high precision
using variational wave functions in Hylleraas coordinates.
The calculations are in accord with quantum-chemical calculations, where available.
Our numerical results  might be useful in constructing accurate three-body potential curves
and for exploration of schemes to create trimers with ultracold atoms in optical lattices using photoassociation.
The formulas listed in the Appendix are general for A($n_0S$)-A($n_0S$)-A($n_0^{\prime}L$), where $L$ is an arbitrary nonzero angular momentum.

\begin{acknowledgments}
This work was supported by NSERC of Canada, the CAS/SAFEA International Partnership Program for Creative Research Teams,
and by the National Basic Research Program of China under Grant No. 2012CB821305 and by NNSF of China under Grant Nos. 11474319, 11104323.
JFB was supported in part by the U. S. NSF through a grant for the Institute of Theoretical Atomic, Molecular, and Optical Physics at
Harvard University and Smithsonian Astrophysical Observatory.
We are grateful to Dr. Richard Schmidt and Dr. Jun-Yi Zhang for helpful discussions.
ZCY thanks Dr. Mang Feng for helpful discussions.
\end{acknowledgments}

\bibliography{positron}

\newpage
\begin{table*}
\caption{Values of $\mathbb{D}_0(M=0)$, $\mathbb{D}_0(M=\pm 1)$, $\mathbb{D}_1(M=0)$, $\mathbb{D}_1(M=\pm1)$, $\mathbb{D}_2$, $\mathbb{Q}_1(M=0)$, $\mathbb{Q}_1(M=\pm1)$, $\mathbb{Q}_2$, $\mathbb{Q}_3(M=0)$, and $\mathbb{Q}_3(M=\pm1)$ for the Li($2\,^2S$)-Li($2\,^2S$)-Li($2\,^{2}P$) system, in atomic units. All these quantities are independent of the geometrical configuration formed by the three atoms. The numbers in parentheses represent the computational uncertainties.}\label{TabI}
\begin{ruledtabular}
\begin{tabular}{lcccccc}
\multicolumn{1}{c}{Atom} & \multicolumn{1}{c}{$\mathbb{D}_0(M=0)$} & \multicolumn{1}{c}{$\mathbb{D}_0(M=\pm 1)$} & \multicolumn{1}{c}{$\mathbb{D}_1(M=0)$}  & \multicolumn{1}{c}{$\mathbb{D}_1(M=\pm1)$} & \multicolumn{1}{c}{$\mathbb{D}_2$} \\
\multicolumn{1}{c}{} & \multicolumn{1}{c}{$\mathbb{Q}_1(M=0)$} & \multicolumn{1}{c}{$\mathbb{Q}_1(M=\pm 1)$} & \multicolumn{1}{c}{$\mathbb{Q}_2$}  & \multicolumn{1}{c}{$\mathbb{Q}_3(M=0)$} & \multicolumn{1}{c}{$\mathbb{Q}_3(M=\pm 1)$} \\
\hline
\\
\multicolumn{1}{c}{$^{\infty}$Li} & $-$5.500111(1) &2.750054(1) &1406.68(3) &1741.06(5) &1393.42(5) \\
\multicolumn{1}{c}{} & 75804.1(5) &354147(4) &83429(1) &27239.28(1) &$-$165583.70(1)\\
\\
\multicolumn{1}{c}{$^{7}$Li}      & $-$5.500926(1) &2.750462(1) &1407.15(5) &1741.59(4) &1394.05(5) \\
\multicolumn{1}{c}{}      & 75809.6(5) &354206(4) &83456(5) &27242.91(2) &$-$165615.09(1)\\
\\
\multicolumn{1}{c}{$^{6}$Li}      & $-$5.501062(1) &2.750530(1) &1407.20(2) &1741.68(4) &1394.16(5) \\
\multicolumn{1}{c}{}      & 75810.5(5) &354216(4) &83460(5) &27243.52(2) &$-$165620.32(1)\\
\end{tabular}
\end{ruledtabular}
\end{table*}


\begingroup
\squeezetable
\begin{table*}
\caption{The additive long-range coefficients $C_3^{(IJ)}(1,M)$ of the Li($2\,^2S$)-Li($2\,^2S$)-Li($2\,^{2}P$) system
for three different types of the zeroth-order wave functions, where the three atoms form an equilateral triangle,
in atomic units. The numbers in parentheses represent the computational uncertainties. }\label{TabII}
\begin{ruledtabular}
\begin{tabular}{lccccccc}
\multicolumn{1}{c}{Atom}  & \multicolumn{1}{c}{State}  & \multicolumn{1}{c}{$C_3^{(12)}(1,M=0)$}  & \multicolumn{1}{c}{$C_3^{(23)}(1,M=0)$} & \multicolumn{1}{c}{$C_3^{(31)}(1,M=0)$}
& \multicolumn{1}{c}{$C_3^{(12)}(1,M=\pm1)$} & \multicolumn{1}{c}{$C_3^{(23)}(1,M=\pm1)$}  & \multicolumn{1}{c}{$C_3^{(31)}(1,M=\pm1)$}\\
\hline
\\
\multicolumn{1}{c}{$^{\infty}$Li}  & $\Psi_{1,\Delta}^{(0)}$   &$-$3.6667415(5)  &$-$3.6667415(5)  &$-$3.6667415(5)  &1.8333702(3)     &1.8333702(3)    &1.8333702(3)\\
\multicolumn{1}{c}{}  &$\Psi_{2,\Delta}^{(0)}$   &0                &0                &5.500110(1)      &0                &0               &$-$2.750054(1)\\
\multicolumn{1}{c}{}  &$\Psi_{3,\Delta}^{(0)}$   &3.6667415(5)     &3.6667415(5)     &$-$1.8333702(3)  &$-$1.8333702(3)  &$-$1.8333702(3) &0.9166850(2)\\
\\
\multicolumn{1}{c}{$^{7}$Li}  &$\Psi_{1,\Delta}^{(0)}$   &$-$3.667284(1)   &$-$3.667284(1)   &$-$3.667284(1)   &1.8336420(5)     &1.8336420(5)     &1.8336420(5)\\
\multicolumn{1}{c}{}  &$\Psi_{2,\Delta}^{(0)}$   &0                &0                &5.500925(2)      &0                &0                &$-$2.750462(1)\\
\multicolumn{1}{c}{}  &$\Psi_{3,\Delta}^{(0)}$   &3.667284(1)      &3.667284(1)      &$-$1.8336420(5)  &$-$1.8336420(5)  &$-$1.8336420(5)  &0.9168210(2)\\
\\
\multicolumn{1}{c}{$^{6}$Li}  &$\Psi_{1,\Delta}^{(0)}$   &$-$3.667374(1)   &$-$3.667374(1)   &$-$3.667374(1)   &1.833686(1)      &1.833686(1)      &1.833686(1)\\
\multicolumn{1}{c}{}  &$\Psi_{2,\Delta}^{(0)}$   &0                &0                &5.501062(1)      &0                &0                &$-$2.750530(1)\\
\multicolumn{1}{c}{}  & $\Psi_{3,\Delta}^{(0)}$   &3.667374(1)      &3.667374(1)      &$-$1.833686(1)   &$-$1.833686(1)   &$-$1.833686(1)   &0.9168437(2)\\
\end{tabular}
\end{ruledtabular}
\end{table*}
\endgroup

\begingroup
\squeezetable
\begin{table*}
\caption{The additive and nonadditive dispersion coefficients $C_6^{(IJ)}(1,M=0)$ and $C_6^{(IJ,JK)}(1,M=0)$ of the Li($2\,^2S$)-Li($2\,^2S$)-Li($2\,^{2}P$) system for three different types of the zeroth-order wave functions, where the three atoms form an equilateral triangle, in atomic units. The numbers in parentheses represent the computational uncertainties.}\label{TabIII}
\begin{ruledtabular}
\begin{tabular}{lccccccc}
\multicolumn{1}{c}{Atom}  &\multicolumn{1}{c}{State}  & \multicolumn{1}{c}{$C_6^{(12)}(1,M=0)$}  & \multicolumn{1}{c}{$C_6^{(23)}(1,M=0)$}  & \multicolumn{1}{c}{$C_6^{(31)}(1,M=0)$} & \multicolumn{1}{c}{$C_6^{(12,23)}(1,M=0)$}  & \multicolumn{1}{c}{$C_6^{(23,31)}(1,M=0)$}  & \multicolumn{1}{c}{$C_6^{(31,12)}(1,M=0)$}\\
\hline
\\
\multicolumn{1}{c}{$^{\infty}$Li}  &$\Psi_{1,\Delta}^{(0)}$  &1402.26(4) &1402.26(4) &1402.26(4)  &157.059(5)     &157.059(5)    &157.059(5)\\
\multicolumn{1}{c}{}  &$\Psi_{2,\Delta}^{(0)}$  &1400.05(4) &1400.05(4) &1406.69(4)  &$-$235.588(7)  &0             &0\\
\multicolumn{1}{c}{}  &$\Psi_{3,\Delta}^{(0)}$  &1404.46(3) &1404.46(3) &1397.85(5)  &78.530(3)      &$-$157.059(5) &$-$157.059(5)\\
\\
\multicolumn{1}{c}{$^{7}$Li}  &$\Psi_{1,\Delta}^{(0)}$  &1402.77(4)  &1402.77(4)  &1402.77(4)  &157.132(5)     &157.132(5)    &157.132(5)\\
\multicolumn{1}{c}{}  &$\Psi_{2,\Delta}^{(0)}$  &1400.59(4)  &1400.59(4)  &1407.14(4)  &$-$235.697(7)  &0             &0\\
\multicolumn{1}{c}{}  &$\Psi_{3,\Delta}^{(0)}$  &1404.95(4)  &1404.95(4)  &1398.41(4)  &78.565(2)      &$-$157.132(5) &$-$157.132(5)\\
\\
\multicolumn{1}{c}{$^{6}$Li}  &$\Psi_{1,\Delta}^{(0)}$  &1402.86(4)  &1402.86(4)  &1402.86(4)  &157.144(5)     &157.144(5)    &157.144(5)\\
\multicolumn{1}{c}{}  &$\Psi_{2,\Delta}^{(0)}$  &1400.68(4)  &1400.68(4)  &1407.20(3)  &$-$235.715(7)  &0             &0\\
\multicolumn{1}{c}{}  &$\Psi_{3,\Delta}^{(0)}$  &1405.02(3)  &1405.02(3)  &1398.50(4)  &78.571(2)      &$-$157.144(5) &$-$157.144(5)\\
\end{tabular}
\end{ruledtabular}
\end{table*}
\endgroup

\begingroup
\squeezetable
\begin{table*}
\caption{The additive and nonadditive dispersion coefficients $C_6^{(IJ)}(1,M=\pm 1)$ and $C_6^{(IJ,JK)}(1,M=\pm 1)$
of the Li($2\,^2S$)-Li($2\,^2S$)-Li($2\,^{2}P$) system
for three different types of the zeroth-order wave functions, where the three atoms form an equilateral triangle, in atomic units.
The numbers in parentheses represent the computational uncertainties.}\label{TabIV}
\begin{ruledtabular}
\begin{tabular}{lccccccc}
\multicolumn{1}{c}{Atom}  & \multicolumn{1}{c}{State}  & \multicolumn{1}{c}{$C_6^{(12)}(1,M=\pm 1)$}  & \multicolumn{1}{c}{$C_6^{(23)}(1,M=\pm 1)$}  & \multicolumn{1}{c}{$C_6^{(31)}(1,M=\pm1)$} & \multicolumn{1}{c}{$C_6^{(12,23)}(1,M=\pm1)$}  & \multicolumn{1}{c}{$C_6^{(23,31)}(1,M=\pm1)$}  & \multicolumn{1}{c}{$C_6^{(31,12)}(1,M=\pm1)$}\\
\hline
\\
\multicolumn{1}{c}{$^{\infty}$Li}  & $\Psi_{1,\Delta}^{(0)}$  &1625.18(5) &1625.18(5)  &1625.18(5)  &$-$137.426(4) &$-$137.426(4)  &$-$137.426(4)\\
\multicolumn{1}{c}{}  & $\Psi_{2,\Delta}^{(0)}$  &1567.24(5) &1567.24(5)  &1741.06(5)  &206.141(7)    &0              &0\\
\multicolumn{1}{c}{}  & $\Psi_{3,\Delta}^{(0)}$  &1683.12(5) &1683.12(5)  &1509.30(5)  &$-$68.713(2)  &137.426(4)     &137.426(4)\\
\\
\multicolumn{1}{c}{$^{7}$Li}  & $\Psi_{1,\Delta}^{(0)}$  &1625.74(4) &1625.74(4)  &1625.74(4)  &$-$137.490(4) &$-$137.490(4)  &$-$137.490(4)\\
\multicolumn{1}{c}{}  & $\Psi_{2,\Delta}^{(0)}$  &1567.82(4) &1567.82(4)  &1741.59(4)  &206.235(6)    &0              &0\\
\multicolumn{1}{c}{}  & $\Psi_{3,\Delta}^{(0)}$  &1683.67(4) &1683.67(4)  &1509.89(4)  &$-$68.745(2)  &137.490(4)     &137.490(4)\\
\\
\multicolumn{1}{c}{$^{6}$Li}  & $\Psi_{1,\Delta}^{(0)}$  &1625.84(4)  &1625.84(4)  &1625.84(4)  &$-$137.500(4) &$-$137.500(4) &$-$137.500(4)\\
\multicolumn{1}{c}{}  & $\Psi_{2,\Delta}^{(0)}$  &1567.92(4)  &1567.92(4)  &1741.68(4)  &206.252(7)    &0             &0\\
\multicolumn{1}{c}{}  & $\Psi_{3,\Delta}^{(0)}$  &1683.76(4)  &1683.76(4)  &1509.99(4)  &$-$68.750(2)  &137.500(4)    &137.500(4)\\
\end{tabular}
\end{ruledtabular}
\end{table*}
\endgroup

\begingroup
\squeezetable
\begin{table*}
\caption{The additive and nonadditive dispersion coefficients $C_8^{(IJ)}(1,M=0)$ and $C_8^{(IJ,JK)}(1,M=0)$
of the Li($2\,^2S$)-Li($2\,^2S$)-Li($2\,^{2}P$) system for three different types of the zeroth-order wave functions, where the three atoms form an equilateral triangle, in atomic units.
The numbers in parentheses represent the computational uncertainties.}\label{TabV}
\begin{ruledtabular}
\begin{tabular}{lccccccc}
\multicolumn{1}{c}{Atom}  &\multicolumn{1}{c}{State}  & \multicolumn{1}{c}{$C_8^{(12)}(1,M=0)$}  & \multicolumn{1}{c}{$C_8^{(23)}(1,M=0)$}  & \multicolumn{1}{c}{$C_8^{(31)}(1,M=0)$} & \multicolumn{1}{c}{$C_8^{(12,23)}(1,M=0)$}  & \multicolumn{1}{c}{$C_8^{(23,31)}(1,M=0)$}  & \multicolumn{1}{c}{$C_8^{(31,12)}(1,M=0)$}\\
\hline
\\
\multicolumn{1}{c}{$^{\infty}$Li}  &$\Psi_{1,\Delta}^{(0)}$  &96505.6(9)  &96505.6(9) &96505.6(9)  &9858.985(6)    &9858.985(6)    &9858.985(6)\\
\multicolumn{1}{c}{}  &$\Psi_{2,\Delta}^{(0)}$  &79616.8(9)  &79616.8(9) &48564.8(5)  &$-$14788.47(1) &0              &0\\
\multicolumn{1}{c}{}  &$\Psi_{3,\Delta}^{(0)}$  &58915.7(8)  &58915.7(8) &89968(2)    &4929.492(3)    &$-$9858.985(6) &$-$9858.985(6)\\
\\
\multicolumn{1}{c}{$^{7}$Li}  &$\Psi_{1,\Delta}^{(0)}$  &96519.5(9)  &96519.5(9) &96519.5(9)  &9861.200(6)    &9861.200(6)    &9861.200(6)\\
\multicolumn{1}{c}{}  &$\Psi_{2,\Delta}^{(0)}$  &79631.3(9)  &79631.3(9) &48566.4(2)  &$-$14791.80(1) &0              &0\\
\multicolumn{1}{c}{}  &$\Psi_{3,\Delta}^{(0)}$  &58921.7(8)  &58921.7(8) &89987(2)    &4930.600(3)    &$-$9861.200(6) &$-$9861.200(6)\\
\\
\multicolumn{1}{c}{$^{6}$Li}  &$\Psi_{1,\Delta}^{(0)}$  &96521.8(9)  &96521.8(9) &96521.8(9)  &9861.567(5)    &9861.567(5)    &9861.567(5)\\
\multicolumn{1}{c}{}  &$\Psi_{2,\Delta}^{(0)}$  &79633.7(9)  &79633.7(9) &48566.9(4)  &$-$14792.35(1) &0              &0\\
\multicolumn{1}{c}{}  &$\Psi_{3,\Delta}^{(0)}$  &58922.7(8)  &58922.7(8) &89990(2)    &4930.784(3)    &$-$9861.567(5) &$-$9861.567(5)\\
\end{tabular}
\end{ruledtabular}
\end{table*}
\endgroup

\begingroup
\squeezetable
\begin{table*}
\caption{The additive and nonadditive dispersion coefficients $C_8^{(IJ)}(1,M=\pm 1)$ and $C_8^{(IJ,JK)}(1,M=\pm 1)$
of the Li($2\,^2S$)-Li($2\,^2S$)-Li($2\,^{2}P$) system
for three different types of the zeroth-order wave functions, where the three atoms form an equilateral triangle, in atomic units.
The numbers in parentheses represent the computational uncertainties.}\label{TabVI}
\begin{ruledtabular}
\begin{tabular}{lccccccc}
\multicolumn{1}{c}{Atom}  &\multicolumn{1}{c}{State}  & \multicolumn{1}{c}{$C_8^{(12)}(1,M=\pm1)$}  & \multicolumn{1}{c}{$C_8^{(23)}(1,M=\pm1)$}  & \multicolumn{1}{c}{$C_8^{(31)}(1,M=\pm1)$} & \multicolumn{1}{c}{$C_8^{(12,23)}(1,M=\pm1)$}  & \multicolumn{1}{c}{$C_8^{(23,31)}(1,M=\pm1)$}  & \multicolumn{1}{c}{$C_8^{(31,12)}(1,M=\pm1)$}\\
\hline
\\
\multicolumn{1}{c}{$^{\infty}$Li}  &$\Psi_{1,\Delta}^{(0)}$  &153520(4)  &153520(4) &153520(4) &$-$26496.02(2) &$-$26496.02(2) &$-$26496.02(2)\\
\multicolumn{1}{c}{}  &$\Psi_{2,\Delta}^{(0)}$  &218790(4)  &218790(4) &519731(4) &39744.04(3)    &0              &0\\
\multicolumn{1}{c}{}  &$\Psi_{3,\Delta}^{(0)}$  &419417(4)  &419417(4) &118475(3) &$-$13248.01(1) &26496.02(2)    &26496.02(2)\\
\\
\multicolumn{1}{c}{$^{7}$Li}  &$\Psi_{1,\Delta}^{(0)}$  &153546(4) &153546(4) &153546(4)  &$-$26501.97(2) &$-$26501.97(2) &$-$26501.97(2)\\
\multicolumn{1}{c}{}  &$\Psi_{2,\Delta}^{(0)}$  &218830(3) &218830(3) &519821(4)  &39752.95(2)    &0              &0\\
\multicolumn{1}{c}{}  &$\Psi_{3,\Delta}^{(0)}$  &419491(4) &419491(4) &118500(3)  &$-$13250.98(1) &26501.97(2)    &26501.97(2)\\
\\
\multicolumn{1}{c}{$^{6}$Li}  &$\Psi_{1,\Delta}^{(0)}$  &153549(3) &153549(3) &153549(3) &$-$26502.97(2) &$-$26502.97(2)  &$-$26502.97(2)\\
\multicolumn{1}{c}{}  &$\Psi_{2,\Delta}^{(0)}$  &218838(4) &218838(4) &519838(5) &39754.44(2)    &0               &0\\
\multicolumn{1}{c}{}  &$\Psi_{3,\Delta}^{(0)}$  &419504(4) &419504(4) &118504(3) &$-$13251.48(1) &26502.97(2)     &26502.97(2)\\
\end{tabular}
\end{ruledtabular}
\end{table*}
\endgroup

\begingroup
\squeezetable
\begin{table*}
\caption{The total long-range interaction coefficients of the Li($2\,^2S$)-Li($2\,^2S$)-Li($2\,^{2}P$) system
for three different types of the zeroth-order wave functions, where the three atoms form an equilateral triangle,
in atomic units. The numbers in parentheses represent the computational uncertainties. }\label{TabII-add}
\begin{ruledtabular}
\begin{tabular}{lccccccc}
\multicolumn{1}{c}{Atom}  & \multicolumn{1}{c}{State}  & \multicolumn{1}{c}{$C_3(1,M=0)$}  & \multicolumn{1}{c}{$C_3(1,M=\pm1)$} & \multicolumn{1}{c}{$C_6(1,M=0)$}
& \multicolumn{1}{c}{$C_6(1,M=\pm1)$} & \multicolumn{1}{c}{$C_8(1,M=0)$}  & \multicolumn{1}{c}{$C_8(1,M=\pm1)$}\\
\hline
\\
\multicolumn{1}{c}{$^{\infty}$Li}  & $\Psi_{1,\Delta}^{(0)}$   &$-$11.0002245(15)  &5.5001106(9)  &4677.95(13)  &4463.26(16)     &319093.8(27)    &381071(12)\\
\multicolumn{1}{c}{}  &$\Psi_{2,\Delta}^{(0)}$   & 5.500110(1)               &$-$2.750054(1)                &3971.20(12)      &5081.68(15)                &193009.9(23)               &997055(12)\\
\multicolumn{1}{c}{}  & $\Psi_{3,\Delta}^{(0)}$
&5.5001128(13)     &$-$2.7500554(8)     &3971.18(12)  &5081.67(16)  &193010.9(18) &997053(11)\\
\end{tabular}
\end{ruledtabular}
\end{table*}
\endgroup

\begingroup
\squeezetable
\begin{table*}
\caption{The additive long-range coefficients $C_3^{(IJ)}(1,M)$ of the Li($2\,^2S$)-Li($2\,^2S$)-Li($2\,^{2}P$) system
for three different types of the zeroth-order wave functions, where the three atoms form an isosceles right triangle,
in atomic units. The numbers in parentheses represent the computational uncertainties.}
\label{TabVII}
\begin{ruledtabular}
\begin{tabular}{lccccccc}
\multicolumn{1}{c}{Atom} & \multicolumn{1}{c}{State} & \multicolumn{1}{c}{$C_3^{(12)}(1,M=0)$} & \multicolumn{1}{c}{$C_3^{(23)}(1,M=0)$} & \multicolumn{1}{c}{$C_3^{(31)}(1,M=0)$} & \multicolumn{1}{c}{$C_3^{(12)}(1,M=\pm1)$}   & \multicolumn{1}{c}{$C_3^{(23)}(1,M=\pm1)$}    & \multicolumn{1}{c}{$C_3^{(31)}(1,M=\pm1)$}\\
\hline
\\
\multicolumn{1}{c}{$^{\infty}$Li} & $\Psi_{1,\bot}^{(0)}$  &$-$3.8591328(7)  &$-$3.0911576(6)  &$-$3.8591328(7) &1.9295663(4)      &1.5455789(3)    &1.9295663(4)\\
\multicolumn{1}{c}{} & $\Psi_{2,\bot}^{(0)}$  &3.8591328(7)     &$-$2.4089529(4)  &3.8591328(7)    &$-$1.9295663(4)   &1.2044764(2)    &$-$1.9295663(4)\\
\multicolumn{1}{c}{} & $\Psi_{3,\bot}^{(0)}$  &0                &5.500111(1)      &0               &0                 &$-$2.7500551(6) &0\\
\\
\multicolumn{1}{c}{$^{7}$Li} & $\Psi_{1,\bot}^{(0)}$  &$-$3.8597053(9) &$-$3.0916162(7) &$-$3.8597053(9) &1.9298527(4)      &1.5458080(4)    &1.9298527(4)\\
\multicolumn{1}{c}{} & $\Psi_{2,\bot}^{(0)}$  &3.8597053(9)    &$-$2.4093103(5) &3.8597053(9)    &$-$1.9298527(4)   &1.2046552(2)    &$-$1.9298527(4)\\
\multicolumn{1}{c}{} & $\Psi_{3,\bot}^{(0)}$  &0               &5.500926(1)     &0               &0                 &$-$2.7504631(7) &0\\
\\
\multicolumn{1}{c}{$^{6}$Li} & $\Psi_{1,\bot}^{(0)}$  &$-$3.8598006(9) &$-$3.0916924(8) &$-$3.8598006(9) &1.9299002(5)      &1.5458460(6)     &1.9299002(5)\\
\multicolumn{1}{c}{} & $\Psi_{2,\bot}^{(0)}$  &3.8598006(9)    &$-$2.4093697(6) &3.8598006(9)    &$-$1.9299002(5)   &1.2046848(3)     &$-$1.9299002(5)\\
\multicolumn{1}{c}{} & $\Psi_{3,\bot}^{(0)}$  &0               &5.501062(1)     &0               &0                 &$-$2.7505310(7)  &0\\
\end{tabular}
\end{ruledtabular}
\end{table*}
\endgroup

\begingroup
\squeezetable
\begin{table*}
\caption{The additive and nonadditive dispersion coefficients $C_6^{(IJ)}(1,M=0)$ and $C_6^{(IJ,JK)}(1,M=0)$ of the Li($2\,^2S$)-Li($2\,^2S$)-Li($2\,^{2}P$) system for three different types of the zeroth-order wave functions, where the three atoms form an isosceles right triangle, in atomic units. The numbers in parentheses represent the computational uncertainties.
}\label{TabVIII}
\begin{ruledtabular}
\begin{tabular}{lccccccc}
\multicolumn{1}{c}{Atom}  &\multicolumn{1}{c}{State}  & \multicolumn{1}{c}{$C_6^{(12)}(1,M=0)$}  & \multicolumn{1}{c}{$C_6^{(23)}(1,M=0)$}  & \multicolumn{1}{c}{$C_6^{(31)}(1,M=0)$} & \multicolumn{1}{c}{$C_6^{(12,23)}(1,M=0)$}  & \multicolumn{1}{c}{$C_6^{(23,31)}(1,M=0)$}  & \multicolumn{1}{c}{$C_6^{(31,12)}(1,M=0)$}\\
\hline
\\
\multicolumn{1}{c}{$^{\infty}$Li}  &$\Psi_{1,\bot}^{(0)}$  &1402.96(4) &1400.87(4) &1402.96(4)  &165.300(5)     &165.300(5)     &132.405(4)\\
\multicolumn{1}{c}{}  &$\Psi_{2,\bot}^{(0)}$  &1403.78(4) &1399.22(4) &1403.78(4)  &$-$165.300(5)  &$-$165.300(5)  &103.183(3)\\
\multicolumn{1}{c}{}  &$\Psi_{3,\bot}^{(0)}$  &1400.05(4) &1406.69(4) &1400.05(4)  &0              &0              &$-$235.588(7)\\
\\
\multicolumn{1}{c}{$^{7}$Li}  &$\Psi_{1,\bot}^{(0)}$  &1403.46(4) &1401.40(4) &1403.46(4)  &165.376(5)     &165.376(5)     &132.466(4)\\
\multicolumn{1}{c}{}  &$\Psi_{2,\bot}^{(0)}$  &1404.27(4) &1399.78(4) &1404.27(4)  &$-$165.376(5)  &$-$165.376(5)  &103.231(3)\\
\multicolumn{1}{c}{}  &$\Psi_{3,\bot}^{(0)}$  &1400.59(4) &1407.14(4) &1400.59(4)  &0              &0              &$-$235.697(7)\\
\\
\multicolumn{1}{c}{$^{6}$Li}  &$\Psi_{1,\bot}^{(0)}$  &1403.54(4) &1401.49(4) &1403.54(4)  &165.389(5)     &165.389(5)     &132.476(4)\\
\multicolumn{1}{c}{}  &$\Psi_{2,\bot}^{(0)}$  &1404.35(4) &1399.87(4) &1404.35(4)  &$-$165.389(5)  &$-$165.389(5)  &103.239(3)\\
\multicolumn{1}{c}{}  &$\Psi_{3,\bot}^{(0)}$  &1400.68(4) &1407.20(3) &1400.68(4)  &0              &0              &$-$235.715(7)\\
\end{tabular}
\end{ruledtabular}
\end{table*}
\endgroup

\begingroup
\squeezetable
\begin{table*}
\caption{The additive and nonadditive dispersion coefficients $C_6^{(IJ)}(1,M=\pm 1)$ and $C_6^{(IJ,JK)}(1,M=\pm 1)$
of the Li($2\,^2S$)-Li($2\,^2S$)-Li($2\,^{2}P$) system
for three different types of the zeroth-order wave functions, where the three atoms form an isosceles right triangle, in atomic units.
The numbers in parentheses represent the computational uncertainties.
}\label{TabIX}
\begin{ruledtabular}
\begin{tabular}{lccccccc}
\multicolumn{1}{c}{Atom}  &\multicolumn{1}{c}{State}  & \multicolumn{1}{c}{$C_6^{(12)}(1,M=\pm1)$}  & \multicolumn{1}{c}{$C_6^{(23)}(1,M=\pm1)$}  & \multicolumn{1}{c}{$C_6^{(31)}(1,M=\pm1)$} & \multicolumn{1}{c}{$C_6^{(12,23)}(1,M=\pm1)$}  & \multicolumn{1}{c}{$C_6^{(23,31)}(1,M=\pm1)$}  & \multicolumn{1}{c}{$C_6^{(31,12)}(1,M=\pm1)$}\\
\hline\\
\multicolumn{1}{c}{$^{\infty}$Li}  &$\Psi_{1,\bot}^{(0)}$  &1643.37(5) &1588.80(5) &1643.37(5) &41.324(1)     &41.324(1)     &$-$264.810(8)\\
\multicolumn{1}{c}{}  &$\Psi_{2,\bot}^{(0)}$  &1664.93(5) &1545.68(5) &1664.93(5) &$-$41.324(1)  &$-$41.324(1)  &$-$206.367(6)\\
\multicolumn{1}{c}{}  &$\Psi_{3,\bot}^{(0)}$  &1567.24(5) &1741.06(5) &1567.24(5) &0             &0             &471.18(2)\\
\\
\multicolumn{1}{c}{$^{7}$Li}  &$\Psi_{1,\bot}^{(0)}$  &1643.94(5) &1589.37(4) &1643.94(5) &41.344(2)     &41.344(2)     &$-$264.932(8)\\
\multicolumn{1}{c}{}  &$\Psi_{2,\bot}^{(0)}$  &1665.48(4) &1546.26(4) &1665.48(4) &$-$41.344(2)  &$-$41.344(2)  &$-$206.462(6)\\
\multicolumn{1}{c}{}  &$\Psi_{3,\bot}^{(0)}$  &1567.83(5) &1741.59(4) &1567.83(5) &0             &0             &471.40(2)\\
\\
\multicolumn{1}{c}{$^{6}$Li}  &$\Psi_{1,\bot}^{(0)}$  &1644.02(4) &1589.48(5) &1644.02(4) &41.348(2)     &41.348(2)     &$-$264.954(9)\\
\multicolumn{1}{c}{}  &$\Psi_{2,\bot}^{(0)}$  &1665.59(5) &1546.36(4) &1665.59(5) &$-$41.348(2)  &$-$41.348(2)  &$-$206.478(6)\\
\multicolumn{1}{c}{}  &$\Psi_{3,\bot}^{(0)}$  &1567.93(5) &1741.68(4) &1567.93(5) &0             &0             &471.42(1)\\
\end{tabular}
\end{ruledtabular}
\end{table*}
\endgroup

\begingroup
\squeezetable
\begin{table*}
\caption{The additive and nonadditive dispersion coefficients $C_8^{(IJ)}(1,M=0)$ and $C_8^{(IJ,JK)}(1,M=0)$
of the Li($2\,^2S$)-Li($2\,^2S$)-Li($2\,^{2}P$) system for three different types of the zeroth-order wave functions, where the three atoms form
an isosceles right triangle, in atomic units.
The numbers in parentheses represent the computational uncertainties.
}\label{TabX}
\begin{ruledtabular}
\begin{tabular}{lccccccc}
\multicolumn{1}{c}{Atom}  & \multicolumn{1}{c}{State}  & \multicolumn{1}{c}{$C_8^{(12)}(1,M=0)$}  & \multicolumn{1}{c}{$C_8^{(23)}(1,M=0)$}  & \multicolumn{1}{c}{$C_8^{(31)}(1,M=0)$} & \multicolumn{1}{c}{$C_8^{(12,23)}(1,M=0)$}  & \multicolumn{1}{c}{$C_8^{(23,31)}(1,M=0)$}  & \multicolumn{1}{c}{$C_8^{(31,12)}(1,M=0)$}\\
\hline
\\
\multicolumn{1}{c}{$^{\infty}$Li}  & $\Psi_{1,\bot}^{(0)}$  &97059.4(9)  &94454(2)    &97059.4(9)  &14674.28(1)    &14674.28(1)     &0\\
\multicolumn{1}{c}{}  & $\Psi_{2,\bot}^{(0)}$  &58361.9(9)  &92021(2)    &58361.9(9)  &$-$14674.28(1) &$-$14674.28(1)  &0\\
\multicolumn{1}{c}{}  & $\Psi_{3,\bot}^{(0)}$  &79617(2)    &48564.8(5)  &79617(2)    &0              &0               &0\\
\\
\multicolumn{1}{c}{$^{7}$Li}  & $\Psi_{1,\bot}^{(0)}$  &97072(1)    &94468(1)    &97072(1)    &14677.57(1)    &14677.57(1)     &0\\
\multicolumn{1}{c}{}  & $\Psi_{2,\bot}^{(0)}$  &58368.8(9)  &92038(2)    &58368.8(9)  &$-$14677.57(1) &$-$14677.57(1)  &0\\
\multicolumn{1}{c}{}  & $\Psi_{3,\bot}^{(0)}$  &79631(1)    &48566.7(5)  &79631(1)    &0              &0               &0\\
\\
\multicolumn{1}{c}{$^{6}$Li}  & $\Psi_{1,\bot}^{(0)}$  &97074(1)    &94472(2)    &97074(1)    &14678.12(1)    &14678.12(1)     &0\\
\multicolumn{1}{c}{}  & $\Psi_{2,\bot}^{(0)}$  &58370(1)    &92041(2)    &58370(1)    &$-$14678.12(1) &$-$14678.12(1)  &0\\
\multicolumn{1}{c}{}  & $\Psi_{3,\bot}^{(0)}$  &79633(1)    &48567.0(5)  &79633(1)    &0              &0               &0\\
\end{tabular}
\end{ruledtabular}
\end{table*}
\endgroup

\begingroup
\squeezetable
\begin{table*}
\caption{The additive and nonadditive dispersion coefficients $C_8^{(IJ)}(1,M=\pm 1)$ and $C_8^{(IJ,JK)}(1,M=\pm 1)$
of the Li($2\,^2S$)-Li($2\,^2S$)-Li($2\,^{2}P$) system
for three different types of the zeroth-order wave functions, where the three atoms form an isosceles right, in atomic units.
The numbers in parentheses represent the computational uncertainties.
}\label{TabXI}
\begin{ruledtabular}
\begin{tabular}{lccccccc}
\multicolumn{1}{c}{Atom}  & \multicolumn{1}{c}{State}  & \multicolumn{1}{c}{$C_8^{(12)}(1,M=\pm1)$}  & \multicolumn{1}{c}{$C_8^{(23)}(1,M=\pm1)$}  & \multicolumn{1}{c}{$C_8^{(31)}(1,M=\pm1)$} & \multicolumn{1}{c}{$C_8^{(12,23)}(1,M=\pm1)$}  & \multicolumn{1}{c}{$C_8^{(23,31)}(1,M=\pm1)$}  & \multicolumn{1}{c}{$C_8^{(31,12)}(1,M=\pm1)$}\\
\hline
\\
\multicolumn{1}{c}{$^{\infty}$Li}  & $\Psi_{1,\bot}^{(0)}$  &161893(4)  &142518(4)  &161893(4)  &$-$16508.56(1) &$-$16508.56(1) &0\\
\multicolumn{1}{c}{}  & $\Psi_{2,\bot}^{(0)}$  &411044(4)  &129477(3)  &411044(4)  &16508.56(1)    &16508.56(1)    &0\\
\multicolumn{1}{c}{}  & $\Psi_{3,\bot}^{(0)}$  &218790(4)  &519731(4)  &218790(4)  &0              &0              &0\\
\\
\multicolumn{1}{c}{$^{7}$Li}  & $\Psi_{1,\bot}^{(0)}$  &161920(4)  &142544(4) &161920(4)  &$-$16512.27(1) &$-$16512.27(1) &0\\
\multicolumn{1}{c}{}  & $\Psi_{2,\bot}^{(0)}$  &411116(3)  &129502(3) &411116(3)  &16512.27(1)    &16512.27(1)    &0\\
\multicolumn{1}{c}{}  & $\Psi_{3,\bot}^{(0)}$  &218830(3)  &519821(4) &218830(3)  &0              &0              &0\\
\\
\multicolumn{1}{c}{$^{6}$Li}  & $\Psi_{1,\bot}^{(0)}$  &161924(4)  &142548(4) &161924(4)  &$-$16512.89(1) &$-$16512.89(1) &0\\
\multicolumn{1}{c}{}  & $\Psi_{2,\bot}^{(0)}$  &411130(4)  &129506(3) &411130(4)  &16512.89(1)    &16512.89(1)    &0\\
\multicolumn{1}{c}{}  & $\Psi_{3,\bot}^{(0)}$  &218838(4)  &519838(5) &218838(4)  &0              &0              &0\\
\end{tabular}
\end{ruledtabular}
\end{table*}
\endgroup

\begingroup
\squeezetable
\begin{table*}
\caption{The additive long-range coefficients $C_3^{(IJ)}(1,M)$ of the Li($2\,^2S$)-Li($2\,^2S$)-Li($2\,^{2}P$) system
for three different types of the zeroth-order wave functions, where the three atoms form a straight line,
in atomic units. The numbers in parentheses represent the computational uncertainties.
}\label{TabXII}
\begin{ruledtabular}
\begin{tabular}{lccccccc}
\multicolumn{1}{c}{Atom}  & \multicolumn{1}{c}{State}  & \multicolumn{1}{c}{$C_3^{(12)}(1,M=0)$} & \multicolumn{1}{c}{$C_3^{(23)}(1,M=0)$} & \multicolumn{1}{c}{$C_3^{(31)}(1,M=0)$}  & \multicolumn{1}{c}{$C_3^{(12)}(1,M=\pm1)$}   & \multicolumn{1}{c}{$C_3^{(23)}(1,M=\pm1)$}    & \multicolumn{1}{c}{$C_3^{(31)}(1,M=\pm1)$}\\
\hline
\\
\multicolumn{1}{c}{$^{\infty}$Li}  & $\Psi_{1,\text{--}}^{(0)}$  &$-$3.8853730(7) &$-$2.8714730(6)  &$-$3.8853730(7) &1.9426864(4)     &1.4357365(3)    &1.9426864(4)\\
\multicolumn{1}{c}{}  & $\Psi_{2,\text{--}}^{(0)}$  &3.8853730(7)    &$-$2.6286373(5)  &3.8853730(7)    &$-$1.9426864(4)  &1.3143187(2)    &$-$1.9426864(4)\\
\multicolumn{1}{c}{}  & $\Psi_{3,\text{--}}^{(0)}$  &0               &5.500111(1)      &0               &0                &$-$2.7500551(6) &0\\
\\
\multicolumn{1}{c}{$^{7}$Li}  & $\Psi_{1,\text{--}}^{(0)}$  &$-$3.885949(1)  &$-$2.8718991(7)  &$-$3.885949(1)  &1.9429746(5)     &1.4359496(3)     &1.9429746(5)\\
\multicolumn{1}{c}{}  & $\Psi_{2,\text{--}}^{(0)}$  &3.885949(1)     &$-$2.6290273(6)  &3.885949(1)     &$-$1.9429746(5)  &1.3145136(3)     &$-$1.9429746(5)\\
\multicolumn{1}{c}{}  & $\Psi_{3,\text{--}}^{(0)}$  &0               &5.500926(1)      &0               &0                &$-$2.7504631(7)  &0\\
\\
\multicolumn{1}{c}{$^{6}$Li}  & $\Psi_{1,\text{--}}^{(0)}$  &$-$3.886045(1)  &$-$2.8719700(7)  &$-$3.886045(1)  &1.9430227(4)     &1.4359849(4)     &1.9430227(4)\\
\multicolumn{1}{c}{}  & $\Psi_{2,\text{--}}^{(0)}$  &3.886045(1)     &$-$2.6290922(6)  &3.886045(1)     &$-$1.9430227(4)  &1.3145461(3)     &$-$1.9430227(4)\\
\multicolumn{1}{c}{}  & $\Psi_{3,\text{--}}^{(0)}$  &0               &5.501062(1)      &0               &0                &$-$2.7505310(7)  &0\\
\end{tabular}
\end{ruledtabular}
\end{table*}
\endgroup

\begingroup
\squeezetable
\begin{table*}
\caption{The additive and nonadditive dispersion coefficients $C_6^{(IJ)}(1,M=0)$ and $C_6^{(IJ,JK)}(1,M=0)$ of the Li($2\,^2S$)-Li($2\,^2S$)-Li($2\,^{2}P$) system for three different types of the zeroth-order wave functions, where the three atoms form a
straight line, in atomic units. The numbers in parentheses represent the computational uncertainties.
}\label{TabXIII}
\begin{ruledtabular}
\begin{tabular}{lccccccc}
\multicolumn{1}{c}{Atom}  & \multicolumn{1}{c}{State}  & \multicolumn{1}{c}{$C_6^{(12)}(1,M=0)$}  & \multicolumn{1}{c}{$C_6^{(23)}(1,M=0)$}  & \multicolumn{1}{c}{$C_6^{(31)}(1,M=0)$} & \multicolumn{1}{c}{$C_6^{(12,23)}(1,M=0)$}  & \multicolumn{1}{c}{$C_6^{(23,31)}(1,M=0)$}  & \multicolumn{1}{c}{$C_6^{(31,12)}(1,M=0)$}\\
\hline
\\
\multicolumn{1}{c}{$^{\infty}$Li}  & $\Psi_{1,\text{--}}^{(0)}$  &1403.22(4) &1400.34(4) &1403.22(4)  &166.425(6)     &166.425(6)     &122.995(4)\\
\multicolumn{1}{c}{}  & $\Psi_{2,\text{--}}^{(0)}$  &1403.50(3) &1399.76(4) &1403.50(3)  &$-$166.425(6)  &$-$166.425(6)  &112.594(4)\\
\multicolumn{1}{c}{}  & $\Psi_{3,\text{--}}^{(0)}$  &1400.05(4) &1406.69(4) &1400.05(4)  &0              &0              &$-$235.588(7)\\
\\
\multicolumn{1}{c}{$^{7}$Li}  & $\Psi_{1,\text{--}}^{(0)}$  &1403.72(4) &1400.88(4) &1403.72(4)  &166.502(6)     &166.502(6)     &123.052(4)\\
\multicolumn{1}{c}{}  & $\Psi_{2,\text{--}}^{(0)}$  &1404.01(4) &1400.30(4) &1404.01(4)  &$-$166.502(6)  &$-$166.502(6)  &112.646(4)\\
\multicolumn{1}{c}{}  & $\Psi_{3,\text{--}}^{(0)}$  &1400.59(4) &1407.14(4) &1400.59(4)  &0              &0              &$-$235.697(7)\\
\\
\multicolumn{1}{c}{$^{6}$Li}  & $\Psi_{1,\text{--}}^{(0)}$  &1403.80(4) &1400.97(4) &1403.80(4)  &166.513(5)     &166.513(5)     &123.060(3)\\
\multicolumn{1}{c}{}  & $\Psi_{2,\text{--}}^{(0)}$  &1404.09(4) &1400.39(4) &1404.09(4)  &$-$166.513(5)  &$-$166.513(5)  &112.655(4)\\
\multicolumn{1}{c}{}  & $\Psi_{3,\text{--}}^{(0)}$  &1400.68(4) &1407.20(3) &1400.68(4)  &0              &0              &$-$235.715(7)\\
\end{tabular}
\end{ruledtabular}
\end{table*}
\endgroup

\begingroup
\squeezetable
\begin{table*}
\caption{The additive and nonadditive dispersion coefficients $C_6^{(IJ)}(1,M=\pm 1)$ and $C_6^{(IJ,JK)}(1,M=\pm 1)$
of the Li($2\,^2S$)-Li($2\,^2S$)-Li($2\,^{2}P$) system
for three different types of the zeroth-order wave functions, where the three atoms form a straight line, in atomic units.
The numbers in parentheses represent the computational uncertainties.
}\label{TabXIV}
\begin{ruledtabular}
\begin{tabular}{lccccccc}
\multicolumn{1}{c}{Atom}  & \multicolumn{1}{c}{State}  & \multicolumn{1}{c}{$C_6^{(12)}(1,M=\pm1)$}  & \multicolumn{1}{c}{$C_6^{(23)}(1,M=\pm1)$}  & \multicolumn{1}{c}{$C_6^{(31)}(1,M=\pm1)$} & \multicolumn{1}{c}{$C_6^{(12,23)}(1,M=\pm1)$}  & \multicolumn{1}{c}{$C_6^{(23,31)}(1,M=\pm1)$}  & \multicolumn{1}{c}{$C_6^{(31,12)}(1,M=\pm1)$}\\
\hline
\\
\multicolumn{1}{c}{$^{\infty}$Li}  & $\Psi_{1,\text{--}}^{(0)}$  &1650.30(4)  &1574.90(4) &1650.30(4)  &416.06(2)    &416.06(2)     &307.487(9)\\
\multicolumn{1}{c}{}  & $\Psi_{2,\text{--}}^{(0)}$  &1657.99(5)  &1559.55(4) &1657.99(5)  &$-$416.06(2) &$-$416.06(2)  &281.483(8)\\
\multicolumn{1}{c}{}  & $\Psi_{3,\text{--}}^{(0)}$  &1567.24(5)  &1741.06(5) &1567.24(5)  &0            &0             &$-$588.97(2)\\
\\
\multicolumn{1}{c}{$^{7}$Li}  & $\Psi_{1,\text{--}}^{(0)}$  &1650.88(5) &1575.49(4) &1650.88(5)  &416.25(1)     &416.25(1)     &307.63(1)\\
\multicolumn{1}{c}{}  & $\Psi_{2,\text{--}}^{(0)}$  &1658.54(4) &1560.16(5) &1658.54(4)  &$-$416.25(1)  &$-$416.25(1)  &281.615(9)\\
\multicolumn{1}{c}{}  & $\Psi_{3,\text{--}}^{(0)}$  &1567.83(5) &1741.59(4) &1567.83(5)  &0             &0             &$-$589.24(2)\\
\\
\multicolumn{1}{c}{$^{6}$Li}  & $\Psi_{1,\text{--}}^{(0)}$  &1650.96(4) &1575.60(5) &1650.96(4)  &416.29(2)     &416.29(2)     &307.653(9)\\
\multicolumn{1}{c}{}  & $\Psi_{2,\text{--}}^{(0)}$  &1658.65(5) &1560.24(4) &1658.65(5)  &$-$416.29(2)  &$-$416.29(2)  &281.636(9)\\
\multicolumn{1}{c}{}  & $\Psi_{3,\text{--}}^{(0)}$  &1567.93(5) &1741.68(4) &1567.93(5)  &0             &0             &$-$589.29(2)\\
\end{tabular}
\end{ruledtabular}
\end{table*}
\endgroup

\begingroup
\squeezetable
\begin{table*}
\caption{The additive and nonadditive dispersion coefficients $C_8^{(IJ)}(1,M=0)$ and $C_8^{(IJ,JK)}(1,M=0)$
of the Li($2\,^2S$)-Li($2\,^2S$)-Li($2\,^{2}P$) system for three different types of the zeroth-order wave functions, where the three atoms form
a straight line, in atomic units.
The numbers in parentheses represent the computational uncertainties.
}\label{TabXV}
\begin{ruledtabular}
\begin{tabular}{lccccccc}
\multicolumn{1}{c}{Atom}  & \multicolumn{1}{c}{State}  & \multicolumn{1}{c}{$C_8^{(12)}(1,M=0)$}  & \multicolumn{1}{c}{$C_8^{(23)}(1,M=0)$}  & \multicolumn{1}{c}{$C_8^{(31)}(1,M=0)$} & \multicolumn{1}{c}{$C_8^{(12,23)}(1,M=0)$}  & \multicolumn{1}{c}{$C_8^{(23,31)}(1,M=0)$}  & \multicolumn{1}{c}{$C_8^{(31,12)}(1,M=0)$}\\
\hline
\\
\multicolumn{1}{c}{$^{\infty}$Li}  & $\Psi_{1,\text{--}}^{(0)}$  &97037.1(9)  &93669(1)   &97037.1(9)  &20893.68(2)    &20893.68(2)     &$-$15441.40(1)\\
\multicolumn{1}{c}{}  & $\Psi_{2,\text{--}}^{(0)}$  &58384.2(9)  &92803(1)   &58384.2(9)  &$-$20893.68(2) &$-$20893.68(2)  &$-$14135.55(1)\\
\multicolumn{1}{c}{}  & $\Psi_{3,\text{--}}^{(0)}$  &79617(1)    &48564.8(5) &79617(1)    &0              &0               &29576.95(2)\\
\\
\multicolumn{1}{c}{$^{7}$Li}  & $\Psi_{1,\text{--}}^{(0)}$  &97049.9(9)  &93685(1)   &97049.9(9)  &20898.36(1)    &20898.36(1)     &$-$15444.87(1)\\
\multicolumn{1}{c}{}  & $\Psi_{2,\text{--}}^{(0)}$  &58391.4(9)  &92821(2)   &58391.4(9)  &$-$20898.36(1) &$-$20898.36(1)  &$-$14138.73(2)\\
\multicolumn{1}{c}{}  & $\Psi_{3,\text{--}}^{(0)}$  &79631(1)    &48566.7(5) &79631(1)    &0              &0               &29583.60(2)\\
\\
\multicolumn{1}{c}{$^{6}$Li}  & $\Psi_{1,\text{--}}^{(0)}$  &97052(1)    &93689(2)    &97052(1)    &20899.14(1)    &20899.14(1)     &$-$15445.45(1)\\
\multicolumn{1}{c}{}  & $\Psi_{2,\text{--}}^{(0)}$  &58392.6(9)  &92822(1)    &58392.6(9)  &$-$20899.14(1) &$-$20899.14(1)  &$-$14139.26(2)\\
\multicolumn{1}{c}{}  & $\Psi_{3,\text{--}}^{(0)}$  &79633(1)    &48567.0(5)  &79633(1)    &0              &0               &29584.70(2)\\
\end{tabular}
\end{ruledtabular}
\end{table*}
\endgroup

\clearpage

\begingroup
\squeezetable
\begin{table*}
\caption{The additive and nonadditive dispersion coefficients $C_8^{(IJ)}(1,M=\pm 1)$ and $C_8^{(IJ,JK)}(1,M=\pm 1)$
of the Li($2\,^2S$)-Li($2\,^2S$)-Li($2\,^{2}P$) system
for three different types of the zeroth-order wave functions, where the three atoms form a straight line, in atomic units.
The numbers in parentheses represent the computational uncertainties.
}\label{TabXVI}
\begin{ruledtabular}
\begin{tabular}{lccccccc}
\multicolumn{1}{c}{Atom}  & \multicolumn{1}{c}{State}  & \multicolumn{1}{c}{$C_8^{(12)}(1,M=\pm1)$}  & \multicolumn{1}{c}{$C_8^{(23)}(1,M=\pm1)$}  & \multicolumn{1}{c}{$C_8^{(31)}(1,M=\pm1)$} & \multicolumn{1}{c}{$C_8^{(12,23)}(1,M=\pm1)$}  & \multicolumn{1}{c}{$C_8^{(23,31)}(1,M=\pm1)$}  & \multicolumn{1}{c}{$C_8^{(31,12)}(1,M=\pm1)$}\\
\hline
\\
\multicolumn{1}{c}{$^{\infty}$Li}  & $\Psi_{1,\text{--}}^{(0)}$  &166509(4) &138319(4) &166509(4) &41787.35(3)    &41787.35(3)     &$-$30882.81(2)\\
\multicolumn{1}{c}{}  & $\Psi_{2,\text{--}}^{(0)}$  &406428(4) &133676(3) &406428(4) &$-$41787.35(3) &$-$41787.35(3)  &$-$28271.10(2)\\
\multicolumn{1}{c}{}  & $\Psi_{3,\text{--}}^{(0)}$  &218790(4) &519731(4) &218790(4) &0              &0               &59153.90(3)\\
\\
\multicolumn{1}{c}{$^{7}$Li}  & $\Psi_{1,\text{--}}^{(0)}$  &166537(4) &138343(3) &166537(4) &41796.72(2)    &41796.72(2)     &$-$30889.75(2)\\
\multicolumn{1}{c}{}  & $\Psi_{2,\text{--}}^{(0)}$  &406499(3) &133701(3) &406499(3) &$-$41796.72(2) &$-$41796.72(2)  &$-$28277.45(2)\\
\multicolumn{1}{c}{}  & $\Psi_{3,\text{--}}^{(0)}$  &218830(3) &519821(4) &218830(3) &0              &0               &59167.19(3)\\
\\
\multicolumn{1}{c}{$^{6}$Li}  & $\Psi_{1,\text{--}}^{(0)}$  &166540(3) &138347(3) &166540(3) &41798.29(2)    &41798.29(2)     &$-$30890.90(2)\\
\multicolumn{1}{c}{}  & $\Psi_{2,\text{--}}^{(0)}$  &406513(4) &133707(4) &406513(4) &$-$41798.29(2) &$-$41798.29(2)  &$-$28278.51(2)\\
\multicolumn{1}{c}{}  & $\Psi_{3,\text{--}}^{(0)}$  &218838(4) &519838(5) &218838(4) &0              &0               &59169.40(3)\\
\end{tabular}
\end{ruledtabular}
\end{table*}
\endgroup


\clearpage

\section{appendix}\label{App}

We consider three like atoms with two atoms in identical $S$ states $|\varphi_{n_0}(0)\rangle$ and the third atom in a non-$S$ state $|\varphi_{n_0'}(LM)\rangle$, where $n_0$ and $n_0'$ are the principal quantum numbers, and $L$ and $M$ are the usual angular quantum numbers.
In the following, as discussed in Sec. \ref{TheA}, we use $\boldsymbol{\sigma}$, $\boldsymbol{\rho}$, and $\boldsymbol{\varsigma}$ to represent collectively the coordinates of each atom. The three orthonormalized degenerate eigenvectors of the unperturbed Hamiltonian with the energy eigenvalue $E_{n_0S;n_0S;n_0'L}^{(0)}=2E_{n_0S}^{(0)}+E_{n_0'L}^{(0)}$ are given in Eqs. (\ref{e6a_a})-(\ref{e6c_c}).
The correct zeroth-order wave functions can always be expanded as a linear combination of $\{\phi_1, \phi_2, \phi_3\}$
\begin{equation}\label{psi0}
|\Psi^{(0)}\rangle=a|\phi_1\rangle+b|\phi_2\rangle+c|\phi_3\rangle \,,
\end{equation}
where \textit{a}, \textit{b}, and \textit{c} are the expansion coefficients with their values depending on the geometrical configuration of the three atoms.

\subsection{The first-order energy correction}\label{AppA1}

According to perturbation theory, the first-order energy correction is
\begin{eqnarray}
&&\Delta E^{(1)}=\langle \Psi^{(0)}|V_{123}|\Psi^{(0)}\rangle \nonumber \\
&&=|a|^{2}\langle\phi_1|V_{123}|\phi_1\rangle+|b|^{2}\langle\phi_2|V_{123}|\phi_2\rangle+
|c|^{2}\langle\phi_3|V_{123}|\phi_3\rangle\nonumber\\
&&+(a^{*}b+b^{*}a)\langle\phi_1|V_{123}|\phi_2\rangle+(a^{*}c+c^{*}a)\langle\phi_1|V_{123}|\phi_3\rangle+
(b^{*}c+c^{*}b)\langle\phi_2|V_{123}|\phi_3\rangle \nonumber \\
&&=(a^{*}b+b^*a)\frac{4\pi}{R_{12}^{2L+1}}\frac{(-1)^{L+M}(2L)!P_{2L}{(\cos\theta_{12})}}{(2L+1)^2(L-M)!(L+M)!}  |\langle\varphi_{n_0}(0;\boldsymbol{\sigma})\|T_L(\boldsymbol{\sigma})
\|\varphi_{n_0'}(L;\boldsymbol{\sigma})\rangle|^2
\nonumber\\
&&+(b^{*}c+c^*b)\frac{4\pi}{R_{23}^{2L+1}}\frac{(-1)^{L+M}(2L)!P_{2L}{(\cos\theta_{23})}}{(2L+1)^2(L-M)!(L+M)!} |\langle\varphi_{n_0}(0;\boldsymbol{\rho})\|T_L(\boldsymbol{\rho})
\|\varphi_{n_0'}(L;\boldsymbol{\rho})\rangle|^2
\nonumber\\
&&+(c^{*}a+a^*c)\frac{4\pi}{R_{31}^{2L+1}}\frac{(-1)^{L+M}(2L)!P_{2L}{(\cos\theta_{31})}}{(2L+1)^2(L-M)!(L+M)!} |\langle\varphi_{n_0}(0;\boldsymbol{\varsigma})\|T_L(\boldsymbol{\varsigma})
\|\varphi_{n_0'}(L;\boldsymbol{\varsigma})\rangle|^2\,.
\end{eqnarray}

\subsection{The second-order energy correction}\label{AppA2}

The second-order energy correction is given by
\begin{eqnarray}\label{E2}
\Delta E^{(2)}&=&-\sum_{n_sn_tn_u}\sum_{L_sL_tL_u}\sum_{M_sM_tM_u}
\frac{|\langle\Psi^{(0)}|V_{123}|\chi_{n_s}(L_sM_s;\boldsymbol{\sigma})\chi_{n_t}(L_tM_t;\boldsymbol{\rho})
\chi_{n_u}(L_uM_u;\boldsymbol{\varsigma})\rangle |^{2}}{E_{n_sL_s;n_tL_t;n_uL_u}-E_{n_0S;n_0S;n_0'L}^{(0)}} \nonumber \\
&=&V_{12}^{(2)}+V_{23}^{(2)}+V_{31}^{(2)}+V_{12,23}^{(2)}+V_{23,31}^{(2)}+V_{31,12}^{(2)}\,,
\end{eqnarray}
where $\chi_{n_s}(L_sM_s;\boldsymbol{\sigma})\chi_{n_t}(L_tM_t;\boldsymbol{\rho})
\chi_{n_u}(L_uM_u;\boldsymbol{\varsigma})$ is an intermediate state of the system with the energy eigenvalue $E_{n_sL_s;n_tL_t;n_uL_u}=E_{n_sL_s}+E_{n_tL_t}+E_{n_uL_u}$. It is noted that the above summations should exclude terms
with $E_{n_sL_s;n_tL_t;n_uL_u}=E_{n_0S;n_0S;n_0'L}^{(0)}$.

In this paper, we choose the coordinate system defined in \ref{TheB} and shown in Fig.~\ref{f1}. Thus in the associated Legendre functions, we have all $\cos(\theta_{IJ})=0$ due to $\theta_{IJ}=\pi/2$. Also in $\exp[i(m_I-m_J)\Phi_{IJ}]$ {\it etc.}, $\Phi_{12}=0$, $\Phi_{23}=\pi-\beta$, and $\Phi_{31}=\pi+\alpha$. Then the three additive terms in the second-order energy correction, denoted by $V_{12}^{(2)}$, $V_{23}^{(2)}$, and $V_{31}^{(2)}$, become respectively
\begin{eqnarray}\label{V12-12b}
V_{12}^{(2)}&=&-|a|^2\sum_{n_sn_t}\sum_{L_sL_t l_1l_{1}^{\prime}}\sum_{M_sM_t m_{1}} \frac{16\pi^2}{R_{12}^{2L_{t}+l_1+l_1^{\prime}+2}}\left(
  \begin{array}{ccc}
    L & l_1 & L_{s}\\
    -M & m_1 & M_{s}\\
  \end{array}
\right)
\left(
  \begin{array}{ccc}
    L & l_{1}^{\prime} & L_{s}\\
    -M & m_{1} & M_{s}\\
  \end{array}
\right)P_{L_{t}+l_1}^{M_{t}-m_1}(0)P_{L_{t}+l_1^{\prime}}^{M_{t}-m_1}(0) \nonumber \\
&&
\frac{(L_{t}+l_1-M_{t}+m_1)!(L_{t}+l_1^{\prime}-M_{t}+m_1)!(L_{t},L_{t})^{-1}(l_1,l_1^{\prime})^{-1/2}}
{(L_{t}+M_{t})!(L_{t}-M_{t})![(l_1+m_1)!(l_1-m_1)!(l_1^{\prime}+m_1)!(l_1^{\prime}-m_1)!]^{1/2}} \nonumber \\
&&
\frac{\langle\varphi_{n_0'}(L;\boldsymbol{\sigma})\|T_{l_1}(\boldsymbol{\sigma})\|\chi_{n_s}(L_s;\boldsymbol{\sigma})\rangle^*
\langle\varphi_{n_0'}(L;\boldsymbol{\sigma})\|T_{l_{1}^{\prime}}(\boldsymbol{\sigma})\|\chi_{n_s}(L_s;\boldsymbol{\sigma})\rangle
|\langle\varphi_{n_0}(0;\boldsymbol{\rho})\|T_{L_t}(\boldsymbol{\rho})\|\chi_{n_t}(L_t;\boldsymbol{\rho})\rangle|^{2}}
{E_{n_sL_s}+E_{n_tL_t}-E_{n_0S}^{(0)}-E_{n_0'L}^{(0)}}
 \nonumber \\
&-&|b|^2\sum_{n_sn_t}\sum_{L_sL_tl_2l_{2}^{\prime}}\sum_{M_sM_t m_{2}}\frac{16\pi^2}
{R_{12}^{2L_{s}+l_2+l_2^{\prime}+2}}\left(
  \begin{array}{ccc}
    L & l_2 & L_{t}\\
    -M & m_2 & M_{t}\\
  \end{array}
\right)
\left(
  \begin{array}{ccc}
    L & l_{2}^{\prime} & L_{t}\\
    -M & m_{2} & M_{t}\\
  \end{array}
\right) P_{L_{s}+l_2}^{M_{s}-m_2}(0) P_{L_{s}+l_2^{\prime}}^{M_{s}-m_2}(0) \nonumber \\
&&\frac{(L_{s}+l_2-M_{s}+m_2)!(L_{s}+l_2^{\prime}-M_{s}+m_2)!(L_{s},L_{s})^{-1}(l_2,l_2^{\prime})^{-1/2}}
{(L_{s}+M_{s})!(L_{s}-M_{s})![(l_2+m_2)!(l_2-m_2)!(l_2^{\prime}+m_2)!(l_2^{\prime}-m_2)!]^{1/2}} \nonumber \\
&&
\frac{|\langle\varphi_{n_0}(0;\boldsymbol{\sigma})\|T_{L_s}(\boldsymbol{\sigma})\|\chi_{n_s}(L_s;\boldsymbol{\sigma})\rangle|^{2}
\langle\varphi_{n_0'}(L;\boldsymbol{\rho})\|T_{l_2}(\boldsymbol{\rho})\|\chi_{n_t}(L_t;\boldsymbol{\rho})\rangle^*
\langle\varphi_{n_0'}(L;\boldsymbol{\rho})\|T_{l_{2}^{\prime}}(\boldsymbol{\rho})\|\chi_{n_t}(L_t;\boldsymbol{\rho})\rangle}
{E_{n_sL_s}+E_{n_tL_t}-E_{n_0S}^{(0)}-E_{n_0'L}^{(0)}}\nonumber  \\
&-&|c|^2\sum_{n_sn_t}\sum_{L_sL_t}\sum_{M_sM_t}\frac{16\pi^2}{R_{12}^{2L_{s}+2L_{t}+2}}
\frac{[P_{L_{s}+L_{t}}^{M_{s}+M_{t}}(0)(L_{s}+L_{t}-M_{s}-M_{t})!]^2(L_{s},L_{t})^{-2}}{(L_{s}+M_{s})!(L_{s}-M_{s})!(L_{t}+M_{t})!(L_{t}-M_{t})!
} \nonumber \\
&&
\frac{|\langle\varphi_{n}(0;\boldsymbol{\sigma})\|T_{L_{s}}(\boldsymbol{\sigma})\|\chi_{n_s}(L_s;\boldsymbol{\sigma})\rangle|^{2}
|\langle\varphi_{n}(0;\boldsymbol{\rho})\|T_{L_t}(\boldsymbol{\rho})\|\chi_{n_t}(L_t;\boldsymbol{\rho})\rangle|^{2}}
{E_{n_sL_s}+E_{n_tL_t}-2E_{n_0S}^{(0)}} \nonumber \\
&-& a^{*}b\sum_{n_sn_t}\sum_{L_sL_t l_1l_{2}^{\prime}}\sum_{M_sM_t m_{1}m_{2}^{\prime}}
\frac{16\pi^2(-1)^{L_{s}+l_2^{\prime}-M_{s}-M_{t}}}{R_{12}^{l_1+L_{s}+L_{t}+l_2^{\prime}+2}}\left(
  \begin{array}{ccc}
    L & l_1 & L_{s}\\
    -M & -m_1 & M_{s}\\
  \end{array}
\right)
\left(
  \begin{array}{ccc}
    L & l_{2}^{\prime} & L_{t}\\
    -M & m_{2}^{\prime} & M_{t}\\
  \end{array}
\right)\nonumber  \\
&&\frac{P_{L_{t}+l_{1}}^{M_{t}+m_{1}}(0)P_{L_{s}+l_{2}^{\prime}}^{M_{s}-m_{2}^{\prime}}(0) (L_{t}+l_{1}-M_{t}-m_{1})!(L_{s}+l_2^{\prime}-M_{s}+m_2^{\prime})!{(L_{s},L_{t})^{-1}(l_1,l_2^{\prime})^{-1/2}}}
{[(L_{s}+M_{s})!(L_{s}-M_{s})!(L_{t}+M_{t})!(L_{t}-M_{t})!(l_{1}+m_{1})!(l_{1}-m_{1})!(l_2^{\prime}+m_2^{\prime})!
(l_2^{\prime}-m_2^{\prime})!]^{1/2}} \nonumber \\
&&
\langle\varphi_{n_0'}(L;\boldsymbol{\sigma})\|T_{l_1}(\boldsymbol{\sigma})\|\chi_{n_s}(L_s;\boldsymbol{\sigma})\rangle^*
\langle\varphi_{n_0}(0;\boldsymbol{\rho})\|T_{L_t}(\boldsymbol{\rho})\|\chi_{n_{t}}(L_t;\boldsymbol{\rho})\rangle^*
\nonumber \\
&&
\frac{\langle\varphi_{n_0}(0;\boldsymbol{\sigma})\|T_{L_s}(\boldsymbol{\sigma})\|\chi_{n_s}(L_s;\boldsymbol{\sigma})\rangle
\langle\varphi_{n_0'}(L;\boldsymbol{\rho})\|T_{l_{2}^{\prime}}(\boldsymbol{\rho})\|\chi_{n_t}(L_t;\boldsymbol{\rho})\rangle}
{E_{n_sL_s}+E_{n_tL_t}-E_{n_0S}^{(0)}-E_{n_0'L}^{(0)}}
 \nonumber \\
&-&b^{*}a\sum_{n_sn_t}\sum_{L_sL_t l_1^{\prime}l_{2}}\sum_{M_sM_t m_{1}^{\prime}m_2} \frac{16\pi^2(-1)^{L_{s}+l_2-M_{s}-M_{t}}}{R_{12}^{l_1^{\prime}
+L_{t}+L_{s}+l_2+2}}\left(
  \begin{array}{ccc}
    L & l_1^{\prime} & L_{s}\\
    -M & -m_1^{\prime} & M_{s}\\
  \end{array}
\right)
\left(
  \begin{array}{ccc}
    L & l_{2} & L_{t}\\
    -M & m_{2} & M_{t}\\
  \end{array}
\right) \nonumber \\
&&\frac{P_{L_{s}+l_{2}}^{M_{s}-m_{2}}(0)P_{L_{t}+l_{1}^{\prime}}^{M_{t}+m_{1}^{\prime}}(0)
(L_{t}+l_{1}^{\prime}-M_{t}-m_{1}^{\prime})!(L_{s}+l_2-M_{s}+m_2)!{(L_{s},L_{t})^{-1}(l_1^{\prime},l_2)^{-1/2}}}
{[(L_{s}+M_{s})!
(L_{s}-M_{s})!(L_{t}+M_{t})!(L_{t}-M_{t})!(l_{1}^{\prime}+m_{1}^{\prime})!(l_{1}^{\prime}-m_{1}^{\prime})!(l_2+m_2)!(l_2-m_2)!]^{1/2}}
\nonumber  \\
&&
\langle\varphi_{n_0}(0;\boldsymbol{\sigma})\|T_{L_{s}}(\boldsymbol{\sigma})\|\chi_{n_s}(L_s;\boldsymbol{\sigma})\rangle^*
\langle\varphi_{n_0'}(L;\boldsymbol{\rho})\|T_{l_{2}}(\boldsymbol{\rho})\|\chi_{n_t}(L_t;\boldsymbol{\rho})\rangle^*
\nonumber \\
&&
\frac{\langle\varphi_{n_0'}(L;\boldsymbol{\sigma})\|T_{l_{1}^{\prime}}(\boldsymbol{\sigma})\|\chi_{n_s}(L_s;\boldsymbol{\sigma})\rangle\langle\varphi_{n_0}(0;\boldsymbol{\rho})\|T_{L_t}(\boldsymbol{\rho})\|\chi_{n_t}(L_t;\boldsymbol{\rho})\rangle
}
{E_{n_sL_s}+E_{n_tL_t}-E_{n_0S}^{(0)}-E_{n_0'L}^{(0)}} \nonumber \\
&=&-\bigg\{|a|^2\sum_{n_sn_t}\sum_{L_sL_t l_1l_{1}^{\prime}} \frac{F_1(n_s,n_t,L_s,L_t;l_1,l_1';L,M)}{R_{12}^{2L_{t}+l_1+l_1^{\prime}+2}}
 \nonumber \\
&+&|b|^2\sum_{n_sn_t}\sum_{L_sL_tl_2l_{2}^{\prime}}
\frac{F_1(n_t,n_s,L_t,L_s;l_2,l_2';L,M)}
{R_{12}^{2L_{s}+l_2+l_2^{\prime}+2}} \nonumber  \\
&+&|c|^2\sum_{n_sn_t}\sum_{L_sL_t}
\frac{F_2(n_s,n_t,L_s,L_t)}{R_{12}^{2L_{s}+2L_{t}+2}}\nonumber \\
&+& a^{*}b\sum_{n_sn_t}\sum_{L_sL_t l_1l_{2}^{\prime}}
\frac{F_3(n_s,n_t,L_s,L_t;l_1,l_2';L,M)}{R_{12}^{L_{s}+L_{t}+l_1+l_2^{\prime}+2}} \nonumber \\
&+& b^{*}a\sum_{n_sn_t}\sum_{L_sL_t l_1^{\prime}l_{2}}
\frac{ F_3^*(n_s,n_t,L_s,L_t;l_1',l_2;L,M)}{R_{12}^{L_{s}+L_{t}+l_1^{\prime}
+l_2+2}}\bigg\} \,,
\end{eqnarray}
\begin{eqnarray}\label{V23-23b}
V_{23}^{(2)}&=&-|a|^2\sum_{n_tn_u}\sum_{L_tL_u}\sum_{M_tM_u}\frac{16\pi^2}{R_{23}^{2L_{t}+2L_{u}+2}}
\frac{[P_{L_{t}+L_{u}}^{M_{t}+M_{u}}(0)(L_{t}+L_{u}-M_{t}-M_{u})!]^2(L_{t},L_{u})^{-2}}{(L_{t}+M_{t})!(L_{t}-M_{t})!(L_{u}+M_{u})!
(L_{u}-M_{u})!} \nonumber \\
&&
\frac{|\langle\varphi_{n_0}(0;\boldsymbol{\rho})\|T_{L_{t}}(\boldsymbol{\rho})\|\chi_{n_t}(L_t;\boldsymbol{\rho})\rangle|^{2}
|\langle\varphi_{n_0}(0;\boldsymbol{\varsigma})\|T_{L_u}(\boldsymbol{\varsigma})\|\chi_{n_u}(L_u;\boldsymbol{\varsigma})\rangle|^{2}}
{E_{n_tL_t}+E_{n_uL_u}-2E_{n_0S}^{(0)}} \nonumber \\
&-&|b|^2\sum_{n_tn_u}\sum_{L_tL_u l_2l_{2}^{\prime}}\sum_{M_tM_u m_2}
\frac{16\pi^2}{R_{23}^{2L_{u}+l_2+l_2^{\prime}+2}}\left(
  \begin{array}{ccc}
    L & l_2 & L_{t}\\
    -M & m_2 & M_{t}\\
  \end{array}
\right)
\left(
  \begin{array}{ccc}
    L & l_{2}^{\prime} & L_{t}\\
    -M & m_{2} & M_{t}\\
  \end{array}
\right)  P_{L_{u}+l_2}^{M_{u}-m_2}(0)P_{{L_{u}+l_2^\prime}}^{M_{u}-m_2}(0) \nonumber \\
&&\frac{(L_{u}+l_2-M_{u}+m_2)!(L_{u}+l_2^{\prime}-M_{u}+m_2)!(L_{u},L_{u})^{-1}(l_2,l_2^{\prime})^{-1/2}}{(L_{u}+M_{u})!
(L_{u}-M_{u})![(l_2+m_2)!(l_2-m_2)!(l_2^{\prime}+m_2)!(l_2^{\prime}-m_2)!]^{1/2}} \nonumber \\
&&
\frac{\langle\varphi_{n_0'}(L;\boldsymbol{\rho})\|T_{l_2}(\boldsymbol{\rho})\|\chi_{n_t}(L_t;\boldsymbol{\rho})\rangle^*
\langle\varphi_{n_0'}(L;\boldsymbol{\rho})\|T_{l_2'}(\boldsymbol{\rho})\|\chi_{n_t}(L_t;\boldsymbol{\rho})\rangle
|\langle\varphi_{n_0}(0;\boldsymbol{\varsigma})\|T_{L_{u}}(\boldsymbol{\varsigma})\|\chi_{n_u}(L_u;\boldsymbol{\varsigma})\rangle|^{2}}
{E_{n_tL_t}+E_{n_uL_u}-E_{n_0S}^{(0)}-E_{n_0'L}^{(0)}} \nonumber \\
&-&|c|^2\sum_{n_tn_u}\sum_{L_tL_u l_3l_{3}^{\prime}}\sum_{M_tM_u m_{3}}
\frac{16\pi^2}{R_{23}^{2L_{t}+l_3+l_3^{\prime}+2}}\left(
  \begin{array}{ccc}
    L & l_3 & L_{u}\\
    -M & m_3 & M_{u}\\
  \end{array}
\right)
\left(
  \begin{array}{ccc}
    L & l_{3}^{\prime} & L_{u}\\
    -M & m_{3} & M_{u}\\
  \end{array}
\right) P_{L_{t}+l_3}^{M_{t}-m_3}(0)P_{L_{t}+l_3^{\prime}}^{M_{t}-m_3}(0) \nonumber \\
&&
\frac{(L_{t}+l_3-M_{t}+m_3)!(L_{t}+l_3^{\prime}-M_{t}+m_3)!(L_{t},L_{t})^{-1}(l_3,l_3^{\prime})^{-1/2}
}{(L_{t}+M_{t})!(L_{t}-M_{t})![(l_3+m_3)!(l_3-m_3)!(l_3^{\prime}+m_3)!(l_3^{\prime}-m_3)!]^{1/2}} \nonumber \\
&&
\frac{|\langle\varphi_{n_0}(0;\boldsymbol{\rho})\|T_{L_{t}}(\boldsymbol{\rho})\|\chi_{n_t}(L_t;\boldsymbol{\rho})\rangle|^{2}
\langle\varphi_{n_0'}(L;\boldsymbol{\varsigma})\|T_{l_3}(\boldsymbol{\varsigma})\|\chi_{n_u}(L_u;\boldsymbol{\varsigma})\rangle^*
\langle\varphi_{n_0'}(L;\boldsymbol{\varsigma})\|T_{l_{3}^{\prime}}(\boldsymbol{\varsigma})\|\chi_{n_u}(L_u;\boldsymbol{\varsigma})\rangle}
{E_{n_tL_t}+E_{n_uL_u}-E_{n_0S}^{(0)}-E_{n_0'L}^{(0)}} \nonumber \\
&-&b^{*}c\sum_{n_tn_u}\sum_{L_tL_u l_2l_{3}^{\prime}}\sum_{M_tM_u m_{2}m_{3}^{\prime}}
\frac{16\pi^2(-1)^{L_{t}+l_3^{\prime}-M_{t}-M_{u}}}{R_{23}^{l_2+L_{u}+L_{t}+l_3^{\prime}+2}}\left(
  \begin{array}{ccc}
    L & l_2 & L_{t}\\
    -M & -m_2 & M_{t}\\
  \end{array}
\right)
\left(
  \begin{array}{ccc}
    L & l_{3}^{\prime} & L_{u}\\
    -M & m_{3}^{\prime} & M_{u}\\
  \end{array}
\right) \nonumber \\
&&\frac{P_{L_{u}+l_{2}}^{M_{u}+m_{2}}(0)P_{L_{t}+l_{3}^{\prime}}^{M_{t}-m_{3}^{\prime}}(0)
(L_{u}+l_{2}-M_{u}-m_{2})!(L_{t}+l_3^{\prime}-M_{t}+m_3^{\prime})!{(L_{t},L_{u})^{-1}(l_2,l_3^{\prime})^{-1/2}}}
{[(l_{2}+m_{2})!(l_{2}-m_{2})!(L_{u}+M_{u})!(L_{u}-M_{u})!(L_{t}+M_{t})!(L_{t}-M_{t})!(l_3^{\prime}+m_3^{\prime})!
(l_3^{\prime}-m_3^{\prime})!]^{1/2}} \nonumber \\
&&
\langle\varphi_{n_0'}(L;\boldsymbol{\rho})\|T_{l_2}(\boldsymbol{\rho})\|\chi_{n_t}(L_t;\boldsymbol{\rho})\rangle^*
\langle\varphi_{n_0}(0;\boldsymbol{\varsigma})\|T_{L_{u}}(\boldsymbol{\varsigma})\|\chi_{n_u}(L_u;\boldsymbol{\varsigma})\rangle^*
\nonumber \\
&&
\frac{\langle\varphi_{n_0}(0;\boldsymbol{\rho})\|T_{L_{t}}(\boldsymbol{\rho})\|\chi_{n_t}(L_t;\boldsymbol{\rho})\rangle
\langle\varphi_{n_0'}(L;\boldsymbol{\varsigma})\|T_{l_{3}^{\prime}}(\boldsymbol{\varsigma})\|\chi_{n_u}(L_u;\boldsymbol{\varsigma})\rangle}
{E_{n_tL_t}+E_{n_uL_u}-E_{n_0S}^{(0)}-E_{n_0'L}^{(0)}}  \nonumber \\
&-& c^{*}b\sum_{n_tn_u}\sum_{L_tL_u l_2^{\prime}l_{3}}\sum_{M_tM_u m_{2}^{\prime}m_{3}}
\frac{16\pi^2(-1)^{L_{t}+l_3-M_{t}-M_{u}}}{R_{23}^{l_2^{\prime}+L_{t}+L_{u}+l_3+2}}\left(
  \begin{array}{ccc}
    L & l_2^{\prime} & L_{t}\\
    -M & -m_2^{\prime} & M_{t}\\
  \end{array}
\right)
\left(
  \begin{array}{ccc}
    L & l_{3} & L_{u}\\
    -M & m_{3} & M_{u}\\
  \end{array}
\right)\nonumber \\
&&\frac{P_{L_{t}+l_{3}}^{M_{t}-m_{3}}(0)P_{L_{u}+l_{2}^{\prime}}^{M_{u}+m_{2}^{\prime}}(0)
(L_{u}+l_{2}^{\prime}-M_{u}-m_{2}^{\prime})!(L_{t}+l_3-M_{t}+m_3)!
{(L_{t},L_{u})^{-1}(l_2^{\prime},l_3)^{-1/2}}}{[(l_{2}^{\prime}+m_{2}^{\prime})!(l_{2}^{\prime}-m_{2}^{\prime})!(L_{u}+M_{u})!
(L_{u}-M_{u})!(L_{t}+M_{t})!(L_{t}-M_{t})!(l_3+m_3)!(l_3-m_3)!]^{1/2}} \nonumber \\
&&
\langle\varphi_{n_0}(0;\boldsymbol{\rho})\|T_{L_{t}}(\boldsymbol{\rho})\|\chi_{n_t}(L_t;\boldsymbol{\rho})\rangle^*
\langle\varphi_{n_0'}(L;\boldsymbol{\varsigma})\|T_{l_{3}}(\boldsymbol{\varsigma})\|\chi_{n_u}(L_u;\boldsymbol{\varsigma})\rangle^*
\nonumber \\
&&
\frac{\langle\varphi_{n_0'}(L;\boldsymbol{\rho})\|T_{l_{2}^{\prime}}(\boldsymbol{\rho})\|\chi_{n_t}(L_t;\boldsymbol{\rho})\rangle
\langle\varphi_{n_0}(0;\boldsymbol{\varsigma})\|T_{L_{u}}(\boldsymbol{\varsigma})\|\chi_{n_u}(L_u;\boldsymbol{\varsigma})\rangle}
{E_{n_tL_t}+E_{n_uL_u}-E_{n_0S}^{(0)}-E_{n_0'L}^{(0)}} \nonumber \\
&=&-\bigg\{|a|^2\sum_{n_tn_u}\sum_{L_tL_u}
\frac{F_2(n_t,n_u,L_t,L_u)}{R_{23}^{2L_{t}+2L_u+2}}
 \nonumber \\
&+&|b|^2\sum_{n_tn_u}\sum_{L_tL_u l_2l_{2}^{\prime}}
\frac{F_1(n_t,n_u,L_t,L_u;l_2,l_2';L,M)}{R_{23}^{2L_{u}+l_2+l_2^{\prime}+2}} \nonumber \\
&+&|c|^2\sum_{n_tn_u}\sum_{L_tL_u l_3l_{3}^{\prime}}
\frac{F_1(n_u,n_t,L_u,L_t;l_3,l_3';L,M)}{R_{23}^{2L_{t}+l_3+l_3^{\prime}+2}} \nonumber \\
&+&b^{*}c\sum_{n_tn_u}\sum_{L_tL_u l_2l_{3}^{\prime}}
\frac{F_3(n_t,n_u,L_t,L_u;l_2,l_3';L,M)}{R_{23}^{l_2+L_{u}+L_{t}+l_3^{\prime}+2}}\nonumber \\
&+& c^{*}b\sum_{n_sn_tn_u}\sum_{L_sL_tL_u l_2^{\prime}l_{3}}
\frac{F_3^{*}(n_t,n_u,L_t,L_u;l_2',l_3;L,M)}{R_{23}^{L_{t}+L_{u}+l_2^{\prime}+l_3+2}} \,,
\bigg\}
\end{eqnarray}
\begin{eqnarray}\label{V31-31b}
V_{31}^{(2)}&=&-|a|^2 \sum_{n_sn_u}\sum_{L_sL_u l_1l_{1}^{\prime}}\sum_{M_sM_u m_{1}}
\frac{16\pi^2}{R_{31}^{2L_{u}+l_1+l_1^{\prime}+2}}\left(
  \begin{array}{ccc}
    L & l_1 & L_{s}\\
    -M & m_1 & M_{s}\\
  \end{array}
\right)
\left(
  \begin{array}{ccc}
    L & l_{1}^{\prime} & L_{s}\\
    -M & m_{1} & M_{s}\\
  \end{array}
\right)\nonumber \\
&&
 P_{L_{u}+l_1}^{M_{u}-m_1}(0)P_{L_{u}+l_1^{\prime}}^{M_{u}-m_1}(0)\frac{(L_{u}+l_1-M_{u}+m_1)!
 (L_{u}+l_1^{\prime}-M_{u}+m_1)!(L_{u},L_{u})^{-1}(l_1,l_1^{\prime})^{-1/2}}{(L_{u}+M_{u})!
(L_{u}-M_{u})![(l_1+m_1)!(l_1-m_1)!(l_1^{\prime}+m_1)!(l_1^{\prime}-m_1)!]^{1/2}}\nonumber \\
&&
\frac{\langle\varphi_{n_0'}(L;\boldsymbol{\sigma})\|T_{l_1}(\boldsymbol{\sigma})\|\chi_{n_s}(L_s;\boldsymbol{\sigma})\rangle^*
\langle\varphi_{n_0'}(L;\boldsymbol{\sigma})\|T_{l_{1}^{\prime}}(\boldsymbol{\sigma})\|\chi_{n_s}(L_s;\boldsymbol{\sigma})\rangle
|\langle\varphi_{n_0}(0;\boldsymbol{\varsigma})\|T_{L_{u}}(\boldsymbol{\varsigma})\|\chi_{n_u}(L_u;\boldsymbol{\varsigma})\rangle|^{2}}
{E_{n_sL_s}+E_{n_uL_u}-E_{n_0S}^{(0)}-E_{n_0'L}^{(0)}}
\nonumber \\
&-&|b|^2\sum_{n_sn_u}\sum_{L_sL_u}\sum_{M_sM_u} \frac{16\pi^2}{R_{31}^{2L_{s}+2L_{u}+2}} \frac{[P_{L_{u}+L_{s}}^{M_{u}+M_{s}}(0)(L_{u}+L_{s}-M_{u}-M_{s})!]^2(L_{u},L_{s})^{-2}}{(L_{u}+M_{u})!(L_{u}-M_{u})!
(L_{s}+M_{s})!(L_{s}-M_{s})!} \nonumber \\
&&
\frac{|\langle\varphi_{n_0}(0;\boldsymbol{\sigma})\|T_{L_{s}}(\boldsymbol{\sigma})\|\chi_{n_s}(L_s;\boldsymbol{\sigma})\rangle|^{2}
|\langle\varphi_{n_0}(0;\boldsymbol{\varsigma})\|T_{L_{u}}(\boldsymbol{\varsigma})\|\chi_{n_u}(L_u;\boldsymbol{\varsigma})\rangle|^{2}}
{E_{n_sL_s}+E_{n_uL_u}-2E_{n_0S}^{(0)}}  \nonumber \\
&-& |c|^2 \sum_{n_sn_u}\sum_{L_sL_u l_3l_{3}^{\prime}}\sum_{M_sM_u m_{3}}
\frac{16\pi^2}{R_{31}^{2L_{s}+l_3+l_3^{\prime}+2}}\left(
  \begin{array}{ccc}
    L & l_3 & L_{u}\\
    -M & m_3 & M_{u}\\
  \end{array}
\right)
\left(
  \begin{array}{ccc}
    L & l_{3}^{\prime} & L_{u}\\
    -M & m_{3} & M_{u}\\
  \end{array}
\right) P_{L_{s}+l_3}^{M_{s}-m_3}(0)P_{L_{s}+l_3^{\prime}}^{M_{s}-m_3}(0) \nonumber \\
&&
\frac{(L_{s}+l_3-M_{s}+m_3)!(L_{s}+l_3^{\prime}-M_{s}+m_3)!(L_{s},L_{s})^{-1}(l_3,l_3^{\prime})^{-1/2}}
{(L_{s}+M_{s})!(L_{s}-M_{s})![(l_3+m_3)!(l_3-m_3)!(l_3^{\prime}+m_3)!(l_3^{\prime}-m_3)!]^{1/2}} \nonumber \\
&&
\frac{|\langle\varphi_{n_0}(0;\boldsymbol{\sigma})\|T_{L_{s}}(\boldsymbol{\sigma})\|\chi_{n_s}(L_s;\boldsymbol{\sigma})\rangle|^{2}
\langle\varphi_{n_0'}(L;\boldsymbol{\varsigma})\|T_{l_3}(\boldsymbol{\varsigma})\|\chi_{n_u}(L_u\boldsymbol{\varsigma})\rangle^*
\langle\varphi_{n_0'}(L;\boldsymbol{\varsigma})\|T_{l_{3}^{\prime}}(\boldsymbol{\varsigma})\|\chi_{n_u}(L_u;\boldsymbol{\varsigma})\rangle}
{E_{n_sL_s}+E_{n_uL_u}-E_{n_0S}^{(0)}-E_{n_0'L}^{(0)}} \nonumber \\
&-& a^{*}c\sum_{n_sn_u}\sum_{L_sL_u l_3^{\prime}l_{1}}\sum_{M_sM_u m_{3}^{\prime}m_{1}}
\frac{16\pi^2(-1)^{L_{u}+l_1-M_{u}-M_{s}}}{R_{31}^{l_3^{\prime}+L_{s}+L_{u}+l_1+2}}\left(
  \begin{array}{ccc}
    L & l_3^{\prime} & L_{u}\\
    -M & -m_3^{\prime} & M_{u}\\
  \end{array}
\right)
\left(
  \begin{array}{ccc}
    L & l_{1} & L_{s}\\
    -M & m_{1} & M_{s}\\
  \end{array}
\right)\nonumber \\
&&\frac{P_{L_{u}+l_{1}}^{M_{u}-m_{1}}(0)P_{L_{s}+l_{3}^{\prime}}^{M_{s}+m_{3}^{\prime}}(0) (L_{s}+l_{3}^{\prime}-M_{s}-m_{3}^{\prime})!(L_{u}+l_1-M_{u}+m_1)!
{(L_{u},L_{s})^{-1}(l_3^{\prime},l_1)^{-1/2}}}{[(L_{s}+M_{s})!(L_{s}-M_{s})!(L_{u}+M_{u})!(L_{u}-M_{u})!
(l_{3}^{\prime}+m_{3}^{\prime})!(l_{3}^{\prime}-m_{3}^{\prime})!
(l_1+m_1)!(l_1-m_1)!]^{1/2}} \nonumber \\
&& \langle\varphi_{n_0'}(L;\boldsymbol{\sigma})\|T_{l_{1}}(\boldsymbol{\sigma})\|\chi_{n_s}(L_s;\boldsymbol{\sigma})\rangle^*
\langle\varphi_{n_0}(0;\boldsymbol{\varsigma})\|T_{L_{u}}(\boldsymbol{\varsigma})\|\chi_{n_u}(L_u;\boldsymbol{\varsigma})\rangle^*
\nonumber \\
&&
\frac{\langle\varphi_{n_0}(0;\boldsymbol{\sigma})\|T_{L_{s}}(\boldsymbol{\sigma})\|\chi_{n_s}(L_s;\boldsymbol{\sigma})\rangle
\langle\varphi_{n_0'}(L;\boldsymbol{\varsigma})\|T_{l_{3}^{\prime}}(\boldsymbol{\varsigma})\|\chi_{n_u}(L_u;\boldsymbol{\varsigma})\rangle}
{E_{n_sL_s}+E_{n_uL_u}-E_{n_0S}^{(0)}-E_{n_0'L}^{(0)}} \nonumber \\
&-&c^{*}a\sum_{n_sn_u}\sum_{L_sL_u l_3l_{1}^{\prime}}\sum_{M_sM_u m_{3}m_{1}^{\prime}}
\frac{16\pi^2(-1)^{L_{u}+l_1^{\prime}-M_{u}-M_{s}}}{R_{31}^{l_3+L_{s}+L_{u}+l_1^{\prime}+2}}\left(
  \begin{array}{ccc}
    L & l_3 & L_{u}\\
    -M & -m_3 & M_{u}\\
  \end{array}
\right)
\left(
  \begin{array}{ccc}
    L & l_{1}^{\prime} & L_{s}\\
    -M & m_{1}^{\prime} & M_{s}\\
  \end{array}
\right)\nonumber \\
&&\frac{P_{L_{s}+l_{3}}^{M_{s}+m_{3}}(0)P_{L_{u}+l_{1}^{\prime}}^{M_{u}-m_{1}^{\prime}}(0)(L_{s}+l_{3}-M_{s}-m_{3})!
(L_{u}+l_1^{\prime}-M_{u}+m_1^{\prime})!{(L_{u},L_{s})^{-1}(l_3,l_1^{\prime})^{-1/2}}}
{[(L_{s}+M_{s})!(L_{s}-M_{s})!(L_{u}+M_{u})!(L_{u}-M_{u})!(l_{3}+m_{3})!(l_{3}-m_{3})!(l_1^{\prime}+m_1^{\prime})!
(l_1^{\prime}-m_1^{\prime})!]^{1/2}} \nonumber \\
&&
\langle\varphi_{n_0'}(L;\boldsymbol{\varsigma})\|T_{l_3}(\boldsymbol{\varsigma})\|\chi_{n_u}(L_u;\boldsymbol{\varsigma})\rangle^*
\langle\varphi_{n_0}(0;\boldsymbol{\sigma})\|T_{L_{s}}(\boldsymbol{\sigma})\|\chi_{n_s}(L_s;\boldsymbol{\sigma})\rangle^*
\nonumber \\
&&
\frac{\langle\varphi_{n_0}(0;\boldsymbol{\varsigma})\|T_{L_u}(\boldsymbol{\varsigma})\|\chi_{n_u}(L_u;\boldsymbol{\varsigma})\rangle
\langle\varphi_{n_0'}(L;\boldsymbol{\sigma})\|T_{l_{1}^{\prime}}(\boldsymbol{\sigma})\|\chi_{n_s}(L_s;\boldsymbol{\sigma})\rangle
}
{E_{n_sL_s}+E_{n_uL_u}-E_{n_0S}^{(0)}-E_{n_0'L}^{(0)}}  \nonumber \\
&=&-\bigg\{
|a|^2 \sum_{n_sn_u}\sum_{L_sL_u l_1l_{1}^{\prime}}
\frac{F_1(n_s,n_u,L_s,L_u;l_1,l_1';L,M)}{R_{31}^{2L_{u}+l_1+l_1^{\prime}+2}}
\nonumber \\
&&+ |b|^2\sum_{n_sn_u}\sum_{L_sL_u} \frac{F_2(n_s,n_u,L_s,L_u)}{R_{31}^{2L_{s}+2L_{u}+2}}\nonumber \\
&&+ |c|^2 \sum_{n_sn_u}\sum_{L_sL_u l_3l_{3}^{\prime}}
\frac{F_1(n_u,n_s,L_u,L_s;l_3,l_3';L,M)}{R_{31}^{2L_{s}+l_3+l_3^{\prime}+2}}
\nonumber \\
&&+ (a^{*}c)\sum_{n_sn_u}\sum_{L_sL_u l_3^{\prime}l_{1}}
\frac{F_3(n_u,n_s,L_u,L_s;l_3',l_1;L,M)}{R_{31}^{l_3^{\prime}+L_{s}+L_{u}+l_1+2}}\nonumber \\
&&+ (c^{*}a)\sum_{n_sn_u}\sum_{L_sL_u l_3l_{1}^{\prime}}
\frac{F_3^*(n_u,n_s,L_u,L_s;l_3,l_1';L,M)}{R_{31}^{l_3+L_{s}+L_{u}+l_1^{\prime}+2}}\bigg\} \,.
\end{eqnarray}
Similarly, the three nonadditive terms are
\begin{eqnarray}\label{V12-23b}
V_{12,23}^{(2)}&=&-\sum_{n_tL_tM_t} \frac{16\pi^2(-1)^{L_{t}+L+M_t-M}}{R_{12}^{L_{t}+L+1}R_{23}^{L_{t}+L+1}} \frac{[P_{L_t+L}^{M_t-M}(0)(L_t +L-M_t+M)!(L_{t},L)^{-1}]^2 }{(L_{t}+M_{t})!(L_{t}-M_{t})!(L+M)!(L-M)!} \nonumber \\
&& \{(a^*c) \exp[-i(M_{t}-M)\beta]+(c^*a) \exp[i(M_{t}-M)\beta]\} \nonumber \\
&&
\frac{|\langle\varphi_{n_0'}(L;\boldsymbol{\sigma})\|T_{L}(\boldsymbol{\sigma})\|\chi_{n_0}(0;\boldsymbol{\sigma})\rangle|^2|\langle\varphi_{n_0}(0;\boldsymbol{\rho})\|T_{L_{t}}(\boldsymbol{\rho})\|\chi_{n_t}(L_t;\boldsymbol{\rho})\rangle|^{2}}
{E_{n_tL_t}-E_{n_0'L}^{(0)}}
\nonumber \\
&-&\sum_{n_tL_tM_t} \frac{16\pi^2(-1)^{L_{t}+L+M_t+M}}{R_{12}^{L_{t}+L+1}R_{23}^{L_{t}+L+1}} \frac{[P_{L_t+L}^{M_t+M}(0)(L_t +L-M_t-M)!(L_{t},L)^{-1}]^2 }{(L_{t}+M_{t})!(L_{t}-M_{t})!(L+M)!(L-M)!} \nonumber \\
&& \{(a^*c) \exp[i(M_{t}+M)\beta]+(c^*a) \exp[-i(M_{t}+M)\beta]\} \nonumber \\
&&
\frac{|\langle\varphi_{n_0}(0;\boldsymbol{\sigma})\|T_{L}(\boldsymbol{\sigma})\|\chi_{n_0'}(L;\boldsymbol{\sigma})\rangle|^2|\langle\varphi_{n_0}(0;\boldsymbol{\rho})\|T_{L_{t}}(\boldsymbol{\rho})\|\chi_{n_t}(L_t;\boldsymbol{\rho})\rangle|^{2}}
{E_{n_tL_t}+E_{n_0'L}-2E_{n_0S}^{(0)}}
\nonumber \\
&=&-\bigg\{\sum_{n_tL_tM_t}\{(a^*c)\exp[-i(M_{t}-M)\beta]\} \frac{ F_4(n_t,L_t,M_t;L,M)
}{R_{12}^{L_{t}+L+1}R_{23}^{L_{t}+L+1}} \nonumber \\
&+&\sum_{n_tL_tM_t} \{(c^*a)\exp[i(M_{t}-M)\beta]\}\frac{ F_4(n_t,L_t,M_t;L,M)
}{R_{12}^{L_{t}+L+1}R_{23}^{L_{t}+L+1}} \bigg\} \,,
\end{eqnarray}
\begin{eqnarray}\label{V23-31b}
V_{23,31}^{(2)}&=&-\sum_{n_uL_uM_u} \frac{16\pi^2(-1)^{L_{u}+L+M_{u}-M}}{R_{23}^{L_{u}+L+1}R_{31}^{L_{u}+L+1}} \frac{[P_{L_u+L}^{M_u-M}(0)(L_u +L-M_u+M)!(L_{u},L)^{-1}]^2 }{(L_{u}+M_{u})!(L_{u}-M_{u})!(L+M)!(L-M)!} \nonumber \\
&& \{(a^*b) \exp[i(M_{u}-M)\gamma]+(b^*a) \exp[-i(M_{u}-M)\gamma]\} \nonumber \\
&&
\frac{|\langle\varphi_{n_0'}(L;\boldsymbol{\sigma})\|T_{L}(\boldsymbol{\sigma})\|\chi_{n_0}(0;\boldsymbol{\sigma})\rangle|^2|\langle\varphi_{n_0}(0;\boldsymbol{\varsigma})\|T_{L_{u}}(\boldsymbol{\varsigma})\|\chi_{n_u}(L_u;\boldsymbol{\varsigma})\rangle|^{2}}
{E_{n_uL_u}-E_{n_0'L}^{(0)}} \nonumber \\
&-&\sum_{n_uL_uM_u} \frac{16\pi^2(-1)^{L_{u}+L+M_{u}+M}}{R_{23}^{L_{u}+L+1}R_{31}^{L_{u}+L+1}} \frac{[P_{L_u+L}^{M_u+M}(0)(L_u +L-M_u-M)!(L_{u},L)^{-1}]^2 }{(L_{u}+M_{u})!(L_{u}-M_{u})!(L+M)!(L-M)!} \nonumber \\
&& \{(a^*b) \exp[-i(M_{u}+M)\gamma]+(b^*a) \exp[i(M_{u}+M)\gamma]\} \nonumber \\
&&
\frac{|\langle\varphi_{n_0}(0;\boldsymbol{\sigma})\|T_{L}(\boldsymbol{\sigma})\|\chi_{n_0'}(L;\boldsymbol{\sigma})\rangle|^2|\langle\varphi_{n_0}(0;\boldsymbol{\varsigma})\|T_{L_{u}}(\boldsymbol{\varsigma})\|\chi_{n_u}(L_u;\boldsymbol{\varsigma})\rangle|^{2}}
{E_{n_uL_u}+E_{n_0'L}-2E_{n_0S}^{(0)}} \nonumber \\
&=& -\bigg\{ \sum_{n_uL_uM_u}\{(a^*b)\exp[i(M_{u}-M)\gamma]\} \frac{F_4(n_u,L_u,M_u;L,M)}{R_{23}^{L_{u}+L+1}R_{31}^{L_{u}+L+1}}\nonumber \\
&+& \sum_{n_uL_uM_u} \{(b^*a)\exp[-i(M_{u}-M)\gamma]\}\frac{F_4(n_u,L_u,M_u;L,M)}{R_{23}^{L_{u}+L+1}R_{31}^{L_{u}+L+1}} \bigg\} \,, \nonumber \\
\end{eqnarray}
\begin{eqnarray}\label{V31-12b}
V_{31,12}^{(2)}&=&-\sum_{n_sL_sM_s} \frac{16\pi^2(-1)^{L_{s}+L+M_{s}-M}}{R_{31}^{L_{s}+L+1}R_{12}^{L_{s}+L+1}} \frac{[P_{L_s+L}^{M_s-M}(0)(L_s +L-M_s+M)!(L_{s},L)^{-1}]^2 }{(L_{s}+M_{s})!(L_{s}-M_{s})!(L+M)!(L-M)!} \nonumber \\
&& \{(b^*c) \exp[i(M_{s}-M)\alpha]+(c^*b) \exp[-i(M_{s}-M)\alpha]\} \nonumber \\
&&\frac{|\langle\varphi_{n_0}(0;\boldsymbol{\sigma})\|T_{L_s}(\boldsymbol{\sigma})\|\chi_{n_s}(L_s;\boldsymbol{\sigma})\rangle|^2|\langle\varphi_{n_0'}(L;\boldsymbol{\rho})\|T_{L}(\boldsymbol{\rho})\|\chi_{n_0}(0;\boldsymbol{\rho})\rangle|^2}
{E_{n_sL_s}-E_{n_0'L}^{(0)}}
 \nonumber \\
&-&\sum_{n_sL_sM_s} \frac{16\pi^2(-1)^{L_{s}+L+M_{s}+M}}{R_{31}^{L_{s}+L+1}R_{12}^{L_{s}+L+1}} \frac{[P_{L_s+L}^{M_s+M}(0)(L_s +L-M_s-M)!(L_{s},L)^{-1}]^2 }{(L_{s}+M_{s})!(L_{s}-M_{s})!(L+M)!(L-M)!} \nonumber \\
&& \{(b^*c) \exp[-i(M_{s}+M)(\alpha)]+(c^*b) \exp[i(M_{s}+M)(\alpha)]\} \nonumber \\
&&\frac{|\langle\varphi_{n_0}(0;\boldsymbol{\sigma})\|T_{L_s}(\boldsymbol{\sigma})\|\chi_{n_s}(L_s;\boldsymbol{\sigma})\rangle|^2|\langle\varphi_{n_0'}(L;\boldsymbol{\rho})\|T_{L}(\boldsymbol{\rho})\|\chi_{n_0}(0;\boldsymbol{\rho})\rangle|^2}
{E_{n_sL_s}+E_{n_0'L}^{(0)}-2E_{n_0S}^{(0)}}
 \nonumber \\
&=&-\bigg\{ \sum_{n_sL_sM_s}\{(b^*c)\exp[i(M_{s}-M)\alpha]\}
\frac{F_4(n_s,L_s,M_s;L,M)}{R_{31}^{L_{s}+L+1}R_{12}^{L_{s}+L+1}}\nonumber \\
&+&\sum_{n_sL_sM_s}\{(c^*b)\exp[-i(M_{s}-M)\alpha]\} \frac{F_4(n_s,L_s,M_s;L,M)}{R_{31}^{L_{s}+L+1}R_{12}^{L_{s}+L+1}} \bigg\} \,.
\end{eqnarray}
In the above, the $F_i$ functions are defined by
\begin{eqnarray}\label{AF1}
F_1(n_s,n_t,L_s,L_t;l_1,l_1';L,M) &=&G_1(L_s,L_t,l_1,l_1';L,M)
\langle\varphi_{n_0'}(L;\boldsymbol{\sigma})\|T_{l_1}(\boldsymbol{\sigma})\|\chi_{n_s}(L_s;\boldsymbol{\sigma})\rangle^*
\nonumber \\
&&\frac{
\langle\varphi_{n_0'}(L;\boldsymbol{\sigma})\|T_{l_{1}^{\prime}}(\boldsymbol{\sigma})\|\chi_{n_s}(L_s;\boldsymbol{\sigma})\rangle
|\langle\varphi_{n_0}(0;\boldsymbol{\rho})\|T_{L_t}(\boldsymbol{\rho})\|\chi_{n_t}(L_t;\boldsymbol{\rho})\rangle|^{2}}
{E_{n_sL_s}+E_{n_tL_t}-E_{n_0S}^{(0)}-E_{n_0'L}^{(0)}} \,, \nonumber \\
\end{eqnarray}
\begin{eqnarray}\label{AF2}
F_2(n_s,n_t,L_s,L_t) &=&G_2(L_s,L_t)
\frac{|\langle\varphi_{n_0}(0;\boldsymbol{\sigma})\|T_{L_{s}}(\boldsymbol{\sigma})\|\chi_{n_s}(L_s;\boldsymbol{\sigma})\rangle|^{2}
|\langle\varphi_{n_0}(0;\boldsymbol{\rho})\|T_{L_t}(\boldsymbol{\rho})\|\chi_{n_t}(L_t;\boldsymbol{\rho})\rangle|^{2}}
{E_{n_sL_s}+E_{n_tL_t}-2E_{n_0S}^{(0)}} \,, \nonumber \\
\end{eqnarray}
\begin{eqnarray}\label{AF3}
F_3(n_s,n_t,L_s,L_t;l_1,l_2';L,M) &=&(-1)^{L_{s}+l_2^{\prime}}G_3(L_s,L_t,l_1,l_2';L,M) \nonumber \\
&&
\langle\varphi_{n_0}(0;\boldsymbol{\sigma})\|T_{L_s}(\boldsymbol{\sigma})\|\chi_{n_s}(L_s;\boldsymbol{\sigma})\rangle^*
\langle\varphi_{n_0'}(L;\boldsymbol{\rho})\|T_{l_{2}^{\prime}}(\boldsymbol{\rho})\|\chi_{n_t}(L_t;\boldsymbol{\rho})\rangle^*
 \nonumber \\
&&
\frac{\langle\varphi_{n_0'}(L;\boldsymbol{\sigma})\|T_{l_1}(\boldsymbol{\sigma})\|\chi_{n_s}(L_s;\boldsymbol{\sigma})\rangle
\langle\varphi_{n_0}(0;\boldsymbol{\rho})\|T_{L_t}(\boldsymbol{\rho})\|\chi_{n_{t}}(L_t;\boldsymbol{\rho})\rangle
}
{E_{n_sL_s}+E_{n_tL_t}-E_{n_0S}^{(0)}-E_{n_0'L}^{(0)}} \,.
 \nonumber \\
\end{eqnarray}
\begin{eqnarray}\label{AF4}
F_4(n_t,L_t,M_t;L,M) &=& (-1)^{L_{t}+L+M_t+M} G_4(L_t,M_t;L,M)
\nonumber \\
&\times&\bigg[
\frac{|\langle\varphi_{n_0'}(L;\boldsymbol{\sigma})\|T_{L}(\boldsymbol{\sigma})\|\chi_{n_0}(0;\boldsymbol{\sigma})\rangle|^2|\langle\varphi_{n_0}(0;\boldsymbol{\rho})\|T_{L_{t}}(\boldsymbol{\rho})\|\chi_{n_t}(L_t;\boldsymbol{\rho})\rangle|^{2}}{E_{n_tL_t}-E_{n_0'L}^{(0)}}
\nonumber \\
&+&
\frac{|\langle\varphi_{n_0'}(L;\boldsymbol{\sigma})\|T_{L}(\boldsymbol{\sigma})\|\chi_{n_0}(0;\boldsymbol{\sigma})\rangle|^2|\langle\varphi_{n_0}(0;\boldsymbol{\rho})\|T_{L_{t}}(\boldsymbol{\rho})\|\chi_{n_t}(L_t;\boldsymbol{\rho})\rangle|^{2}}
{E_{n_tL_t}+E_{n_0'L}^{(0)}-2E_{n_0S}^{(0)}}\bigg] \,, \nonumber \\
\end{eqnarray}
where $G_1(L_i,L_j,\ell_k,\ell_k';L,M)$, $G_2(L_i,L_j)$, $G_3(L_i,L_j,\ell_{k_1},\ell_{k_2}';L,M)$, and $G_4(L_i,M_i;L,M)$ are further defined by:
\begin{eqnarray}\label{G1}
G_1(L_i,L_j,\ell_k,\ell_k';L,M)
&=& \frac{16\pi^2(\ell_k,\ell_k^{\prime})^{-1/2}}{(2L_{j}+1)^2}\sum_{M_i M_j m_k} \left(
  \begin{array}{ccc}
    L & \ell_k & L_i\\
    -M & m_k & M_i\\
  \end{array}
\right)
\left(
  \begin{array}{ccc}
    L & \ell_{k}^{\prime} & L_i\\
    -M & m_{k} & M_i\\
  \end{array}
\right) \nonumber \\
&&
\frac{(L_j+\ell_k-M_j+m_k)!(L_j+\ell_k^{\prime}-M_j+m_k)!P_{L_j+\ell_k}^{M_j-m_k}(0)P_{L_j+\ell_k^{\prime}}^{M_j-m_k}(0)}
{(L_j+M_j)!(L_j-M_j)![(\ell_k+m_k)!(\ell_k-m_k)!(\ell_k^{\prime}+m_k)!(\ell_k^{\prime}-m_k)!]^{1/2}} \,, \nonumber \\ \\
G_2(L_i,L_j)&=& 16\pi^2(L_i,L_j)^{-2}\sum_{M_iM_j}
\frac{[P_{L_i+L_j}^{M_i+M_j}(0)(L_i+L_j-M_i-M_j)!]^2}{(L_i+M_i)!
(L_i-M_i)!(L_j+M_j)!(L_j-M_j)!} \,, \nonumber \\ \\
G_3(L_i,L_j,\ell_{k_1},\ell_{k_2}';L,M)&=&\frac{16\pi^2(\ell_{k_1},\ell_{k_2}^{\prime})^{-1/2}}{(2L_i+1)(2L_j+1)}\sum_{M_iM_j m_{k_{1}}m_{k_2}^{\prime}}
\left(
  \begin{array}{ccc}
    L & \ell_{k_1} & L_{i}\\
    -M & -m_{k_1} & M_{i}\\
  \end{array}
\right)
\left(
  \begin{array}{ccc}
    L & \ell_{k_2}^{\prime} & L_{j}\\
    -M & m_{k_2}^{\prime} & M_{j}\\
  \end{array}
\right) \nonumber \\
&& \frac{(-1)^{M_{i}+M_{j}}P_{L_{j}+\ell_{k_1}}^{M_{j}+m_{k_1}}(0)P_{L_{i}+\ell_{k_2}^{\prime}}^{M_{i}-m_{k_2}^{\prime}}(0)}
{[(L_{i}+M_{i})!(L_{i}-M_{i})!(L_{j}+M_{j})!(L_{j}-M_{j})!]^{1/2}} \nonumber  \\
&&\frac{ (L_{j}+\ell_{k_1}-M_{j}-m_{k_1})!(L_{i}+\ell_{k_2}^{\prime}-M_{i}+m_{k_2}^{\prime})!}
{[(\ell_{k_1}+m_{k_1})!(\ell_{k_1}-m_{k_1})!(\ell_{k_2}^{\prime}+m_2^{\prime})!
(\ell_{k_2}^{\prime}-m_2^{\prime})!]^{1/2}} \, , \nonumber \\ \\
G_4(L_i,M_i;L,M)&=&16\pi^2\frac{[P_{L_i+L}^{M_i-M}(0)(L_i+L-M_i+M)!(L_{i},L)^{-1}]^2 }{(L_{i}+M_{i})!(L_{i}-M_{i})!(L+M)!(L-M)!} \,.
\end{eqnarray}
Then the second-order energy correction is simplified as,
\begin{eqnarray}
\Delta E^{(2)}
&=&-\sum_{n\geq 3}\bigg(\frac{C_{2n}^{(12)}(L,M)}{R_{12}^{2n}}+\frac{C_{2n}^{(23)}(L,M)}{R_{23}^{2n}}+\frac{C_{2n}^{(31)}(L,M)}{R_{31}^{2n}} \nonumber \\
&+&\frac{C_{2n}^{(12,23)}(L,M)}
{R_{12}^nR_{23}^n}+\frac{C_{2n}^{(23,31)}(L,M)}{R_{23}^nR_{31}^n}+\frac{C_{2n}^{(31,12)}(L,M)}{R_{31}^nR_{12}^n}\bigg) \nonumber \\ \,
\end{eqnarray}
where $C_{2n}^{(IJ)}(L,M)$ and $C_{2n}^{(IJ,JK)}(L,M)$ are, respectively, the additive and nonadditive dispersion coefficients. These coefficients can be expressed as
\begin{eqnarray}
C_{2n}^{(12)}(L,M)&=&|a|^2\sum_{n_sn_t}\sum_{\substack{L_sL_t l_1l_{1}^{\prime} \\ 2L_{t}+l_1+l_1^{\prime}+2=2n}}
F_1(n_s,n_t,L_s,L_t;l_1,l_1';L,M) \nonumber \\
&+&|b|^2\sum_{n_sn_t}\sum_{\substack{L_sL_tl_2l_{2}^{\prime} \\ 2L_{s}+l_2+l_2^{\prime}+2=2n}}
F_1(n_t,n_s,L_t,L_s;l_2,l_2';L,M) \nonumber  \\
&+&|c|^2\sum_{n_sn_t} \sum_{\substack{L_sL_t \\ 2L_{s}+2L_{t}+2 =2n}}
F_2(n_s,n_t,L_s,L_t)\nonumber \\
&+& a^{*}b\sum_{n_sn_t}\sum_{\substack{L_sL_t l_1l_{2}^{\prime} \\ L_{s}+L_{t}+l_1+l_2^{\prime}+2=2n}}
F_3(n_s,n_t,L_s,L_t;l_1,l_2';L,M) \nonumber \\
&+& b^{*}a\sum_{n_sn_t}\sum_{\substack{L_sL_t l_1^{\prime}l_{2} \\ L_{s}+L_{t}+l_1^{\prime}
+l_2+2=2n}}F_3^*(n_s,n_t,L_s,L_t;l_1',l_2;L,M) \,,
\end{eqnarray}
\begin{eqnarray}
C_{2n}^{(23)}(L,M)&=&|a|^2\sum_{n_tn_u}\sum_{\substack{L_tL_u  \\ 2L_{t}+2L_u+2 =2n}} F_2(n_t,n_u,L_t,L_u)
 \nonumber \\
&+&|b|^2\sum_{n_tn_u}\sum_{\substack{L_tL_u l_2l_{2}^{\prime} \\ 2L_{u}+l_2+l_2^{\prime}+2 =2n}} F_1(n_t,n_u,L_t,L_u;l_2,l_2';L,M)
 \nonumber \\
&+&|c|^2\sum_{n_tn_u}\sum_{\substack{L_tL_u l_3l_{3}^{\prime} \\ 2L_{t}+l_3+l_3^{\prime}+2=2n}} F_1(n_u,n_t,L_u,L_t;l_3,l_3';L,M)
\nonumber \\
&+&b^{*}c\sum_{n_tn_u}\sum_{\substack{L_tL_u l_2l_{3}^{\prime} \\ L_{u}+L_{t}+l_2+l_3^{\prime}+2 =2n}}
F_3(n_t,n_u,L_t,L_u;l_2,l_3';L,M)\nonumber \\
&+& c^{*}b\sum_{n_tn_u}\sum_{\substack{L_sL_tL_u l_2^{\prime}l_{3} \\ L_{t}+L_{u}+l_2^{\prime}+l_3+2 =2n}}
F_3^*(n_t,n_u,L_t,L_u;l_2',l_3;L,M) \,,
\end{eqnarray}
\begin{eqnarray}
C_{2n}^{(31)}(L,M)&=&|a|^2 \sum_{n_sn_u}\sum_{\substack{L_sL_u l_1l_{1}^{\prime} \\ 2L_{u}+l_1+l_1^{\prime}+2 =2n}}
F_1(n_s,n_u,L_s,L_u;l_1,l_1';L,M)
\nonumber \\
&+& |b|^2\sum_{n_sn_u}\sum_{\substack{L_sL_u  \\ 2L_{s}+2L_{u}+2=2n}} F_2(n_s,n_u,L_s,L_u) \nonumber \\
&+& |c|^2 \sum_{n_sn_u}\sum_{\substack{L_sL_u l_3l_{3}^{\prime} \\ 2L_{s}+l_3+l_3^{\prime}+2 =2n}}
F_1(n_u,n_s,L_u,L_s;l_3,l_3';L,M)
\nonumber \\
&+& (a^{*}c)\sum_{n_sn_u}\sum_{\substack{L L_sL_u l_3^{\prime}l_{1} \\L_{s}+L_{u}+l_1+l_3^{\prime}+2=2n}}
F_3^*(n_u,n_s,L_u,L_s;l_3',l_1;L,M) \nonumber \\
&+& (c^{*}a)\sum_{n_sn_u}\sum_{\substack{L_sL_u l_3l_{1}^{\prime} \\L_{s}+L_{u}+l_1^{\prime}+l_3+2 =2n}}
F_3(n_u,n_s,L_u,L_s;l_3,l_1';L,M) \,,
\end{eqnarray}
\begin{eqnarray}
C_{2n}^{(12,23)}(L,M)&=&\sum_{\substack{n_tL_tM_t  \\ L_{t}+L+1 = n}}
\{(a^*c)\exp[-i(M_{t}-M)\beta]\}F_4(n_t,L_t,M_t;L,M) \nonumber \\
&+& \sum_{\substack{n_tL_tM_t  \\ L_{t}+L+1 = n}}
\{(c^*a)\exp[i(M_{t}-M)\beta]\}F_4(n_t,L_t,M_t;L,M) \,,
\end{eqnarray}
\begin{eqnarray}
C_{2n}^{(23,31)}(L,M)&=&\sum_{\substack{n_uL_uM_u   \\ L_{u}+L+1 = n}}
\{(a^*b)\exp[i(M_{u}-M)\gamma]\} F_4(n_u,L_u,M_u;L,M) \nonumber \\
&+& \sum_{\substack{n_uL_uM_u   \\ L_{u}+L+1 = n}}
\{(b^*a)\exp[-i(M_{u}-M)\gamma]\}F_4(n_u,L_u,M_u;L,M) \,,
\end{eqnarray}
\begin{eqnarray}
C_{2n}^{(31,12)}(L,M)&=&\sum_{\substack{n_sL_sM_s    \\ L_{s}+L+1 = n}}
\{(b^*c)\exp[i(M_{s}-M)\alpha]\}F_4(n_s,L_s,M_s;L,M)\nonumber \\
&+&\sum_{\substack{n_sL_sM_s \\ L_{s}+L+1 = n}}
\{(c^*b)\exp[-i(M_{s}-M)\alpha]\} F_4(n_s,L_s,M_s;L,M) \,.
\end{eqnarray}

\end{document}